\documentclass[12pt]{article}
\usepackage{a4wide}
\usepackage{amsmath,amssymb,bbm}
\usepackage{multirow}
\usepackage{youngtab}
\usepackage{comment}

\def\2{\frac12}
\def\4{\frac14}

\newcommand{\be}{\begin{equation}}
\newcommand{\ee}{\end{equation}}
\newcommand{\bea}{\begin{eqnarray}}
\newcommand{\eea}{\end{eqnarray}}
\def\a{\alpha}
\def\b{\beta}
\def\g{\gamma}

\def\d{\delta}
\def\e{\epsilon}

\def\m{\mu}
\def\n{\nu}

\def\r{\rho}

\begin{document}

\begin{titlepage}
\begin{center}

\hfill UG-10-33 \\ \hfill KCL-MTH-10-11

\vskip 1.5cm

{\Large \bf  D-Brane Wess-Zumino Terms and U-Duality}

\vskip 1cm

{\bf Eric A.~Bergshoeff\,$^1$ and Fabio Riccioni\,$^2$}

\vskip 25pt

{\em $^1$ \hskip -.1truecm Centre for Theoretical Physics,
University of Groningen, \\ Nijenborgh 4, 9747 AG Groningen, The
Netherlands \vskip 5pt }

{email: {\tt E.A.Bergshoeff@rug.nl}} \\

\vskip 15pt

{\em $^2$ \hskip -.1truecm Department of Mathematics, Kings College
London, \\ Strand London, WC2R 2LS UK \vskip 5pt }

{email: {\tt Fabio.Riccioni@kcl.ac.uk}} \\

\end{center}

\vskip 0.5cm

\begin{center} {\bf ABSTRACT}\\[3ex]
\end{center}

We construct gauge-invariant and U-duality covariant expressions for
Wess-Zumino terms corresponding to general D$p$-branes ($0\le p \le
D-1$) in arbitrary $3\le D\le 10$ dimensions. A distinguishing
feature of these Wess-Zumino terms is that they contain twice as
many scalars as the $10-D$ compactified dimensions, in line with
doubled geometry. We find that for $D<10$ the charges of the
higher-dimensional branes can all be expressed as products of the
0-brane charges, which include the D0-brane and the NS-NS 0-brane
charges. We give the general expressions for these charges and show
how they determine the non-trivial conjugacy class to which some of
the higher-dimensional D-branes belong.

\end{titlepage}

\newpage
\setcounter{page}{1} \tableofcontents

\newpage

\setcounter{page}{1} \numberwithin{equation}{section}

\section{Introduction}

Since their invention in 1995 D-branes have played a crucial role in
understanding the (non-perturbative) nature of string theory
\cite{Polchinski:1995mt}. D-branes occur as brane solutions of the
low-energy supergravity limit of string theory
\cite{Duff:1994an}. Their dynamics and interactions
can be described by an appropriate worldvolume action. Unlike the
usual brane actions the D-brane actions do not only contain
worldvolume embedding scalars describing the position of the brane
but also a worldvolume vector field describing the fact that
Fundamental strings may end on the brane. It is well-known that the
kinetic terms for the embedding scalars and worldvolume vector are
given by a so-called Dirac-Born-Infeld (DBI) action. In view of this
the worldvolume vector is often called the Born-Infeld (BI) vector.
The coupling of the D-brane to the Ramond-Ramond (RR) gauge
potentials is described by a so-called Wess-Zumino (WZ)
term.\,\footnote{ A RR field is defined as a field that describes a
(massless) string state created by two fermionic oscillators in the
Ramond sector. Such fields couple to D-branes. Here we extend this
definition to include fields that do not describe physical degrees
of freedom but do couple to D-branes, e.g.~domain walls and
space-filling branes. These fields are related to the
RR-potentials describing physical degrees of freedom via T-duality.}

Using a short-hand notation the WZ terms of all D$p$-branes, with
$0\le p\le9$, in Type II string theory are given by
\cite{Polchinski}\,\footnote{We do not consider higher-order
corrections involving the (target space) Riemann curvature tensor,
like in \cite{Green:1996dd}.}
  \begin{equation}\label{WZterm}
  {\cal L}_{\text{WZ}}\text{(D=10)} = e^{{\cal F}_2}C\,.
  \end{equation}
Here $ C$ is defined as the formal sum\,\footnote{Note that this sum
also contains a term involving the axion field $C_0$ which is not
required by gauge-invariance of the IIB theory. It can be predicted
by combining gauge invariance of the IIA theory and T-duality
\cite{Bergshoeff:1996cy}. In this work we will first concentrate on
gauge invariance and only afterwards, see end of Section 5, consider the dependence of the
Wess-Zumino term on the axionic scalars.}
  \begin{equation}
  C = \sum_n C_n\,,
  \end{equation}
where $C_n\ (1\le n\le 10)$ are the pull-backs of the RR $n$-form
potentials, i.e.
\begin{equation}
C_{i_1 ...i_n} = \partial_{i_1}X^{\mu_1}\cdots
\partial_{i_n}X^{\mu_n}\,C_{\mu_1\cdots\mu_n}\,.
\end{equation}
The $X^\mu\ (\mu=0,1,\cdots,10)$ are the embedding worldvolume
scalars and $\partial_i\ (i=0,1,\cdots ,p)$ is the derivative with
respect to the worldvolume coordinates of the D$p$-brane.
Furthermore, ${\cal F}_2$ is the worldvolume 2-form curvature tensor
of the BI vector $V_1$ given by
  \begin{equation}
  {\cal F}_2 = dV_1 +B_2\,,
  \end{equation}
where $B_2$ is the Neveu-Schwarz (NS-NS) 2-form potential. Using the
above notation and the Bianchi identity
\begin{equation}
d{\cal F}_2=H_3\equiv dB_2
\end{equation}
 one can
show that the WZ term is invariant under
  \begin{equation}\label{specialform}
  \delta C = d\lambda + H_3\wedge \lambda\,,
  \end{equation}
where $\lambda$ is the formal sum of the RR gauge parameters $\lambda_n\ (0\le n\le 9)$:
  \begin{equation}
  \lambda = \sum_n \lambda_n\,.
  \end{equation}

Type IIA string theory has D$p$-branes for all even $p\
(p=0,2,4,6,8)$. The D0-brane is a particle, the D2-brane a membrane
etc. The D8-brane is a domain-wall (one transverse direction) which
couples to the RR 9-form potential \cite{Bergshoeff:1996ui} that is
dual to the Romans parameter $m$ in massive IIA supergravity
\cite{Romans:1985tz}. On the other hand, Type IIB string theory has
D$p$-branes for all odd $p\ (p=1,3,5,7,9)$. The D1-brane is the
D-string etc. The D9-brane is a so-called space-filling brane that
couples to the RR 10-form $C_{10}$ of IIB supergravity. Note that
the 10-form potential does not describe any physical degree of
freedom and is therefore often not given in combination with the
standard IIB supergravity multiplet. Nevertheless, it is needed to
describe the coupling of the D9-brane to IIB supergravity and it is
perfectly consistent with the IIB superalgebra to add this potential
to the standard multiplet \cite{Bergshoeff:2005ac}.

A distinguishing feature of Type IIB string theory is that it has a
manifest $\text{SL}(2,\mathbb{R})$ S-duality under which the
D-branes transform non-trivially. For instance, the D1-brane
transforms, together with the Fundamental string F1, as a doublet of
$\text{SL}(2,\mathbb{R})$. This can  be seen from the fact
that IIB supergravity has a doublet of 2-form potentials
$A^\alpha_{2}\ (\alpha=1,2)$. The two components of the doublet
describe the NS-NS and RR 2-forms. Similarly, the other even-form
potentials of IIB supergravity transform as a singlet, doublet,
triplet and quadruplet+doublet, respectively:
  \begin{equation}\label{A-fields}
  A_2^\alpha\,,\ A_4\,,\ A_6^\alpha\,,\ A_8^{\alpha\beta}\,,\
  A_{10}^{\alpha\beta\gamma}\,,\ A_{10}^\alpha\,.
  \end{equation}
The 10-forms are special in the sense that they occur as a
reducible representation of the duality group: a quadruplet and a
doublet. It turns out that it is the
quadruplet that  contains the RR 10-form that couples to the
D9-brane \cite{Bergshoeff:2005ac}.

In order to write down the WZ terms for the D-branes of Type IIB
string theory  one needs to relate the S-duality covariant
 $A$-fields given in \eqref{A-fields} to the
RR-fields $C$ occurring in the expression \eqref{WZterm} of the WZ
term. In order to do this in a duality-covariant way one first
introduces charge vectors ${\tilde q}_\alpha, q_\alpha$ and defines
\cite{Bergshoeff:2006gs}
  \begin{equation}\label{basicIIBcharges}
 C_2={\tilde q}_\alpha A_2^\alpha\,,\hskip 1.5truecm B_2 = q_\alpha A_2^\alpha\,,
  \end{equation}
with ${\tilde q}_\alpha q_\beta\epsilon^{\alpha\beta}\ne
0$,\,\footnote{ In \cite{Bergshoeff:2006gs} we took ${\tilde
q}_\alpha q_\beta\epsilon^{\alpha\beta}=i.$ In this paper we will
leave the normalisation unfixed so that we can factor out the
charges as a common factor in front of the WZ term.} such that $C_2$
is the RR 2-form coupling to the D1-string and $B_2$ is the NS-NS
2-form coupling to the Fundamental string. Assuming that ${\tilde
q}_\alpha$ and $q_\alpha$ transform as  doublets under
$\text{SL}(2,\mathbb{R})$ this fixes in a duality-covariant way
which component of the 2-form doublet we define as the RR 2-form and
which component as the NS-NS 2-form.

The leading term in the  WZ-term of the D3-brane, i.e.~(the
pull-back of) the RR 4-form can be written as
\cite{Bergshoeff:2006gs}
  \begin{equation}\label{relation}
  C_4 = {\tilde
  q}_\alpha q_\beta\bigl( - i \epsilon^{\alpha\beta}A_4 -\tfrac{1}{16}
  A_2^\alpha A_2^\beta\bigr)\,.
  \end{equation}
The difference between $A_4$ and $C_4$ is that $A_4$ transforms
under both the RR gauge transformations of $C_2,C_4$ and the NS-NS
gauge transformations of $B_2$ whereas $C_4$ only transforms under
the RR gauge transformations. The requirement that $C_4$ transforms
as in \eqref{specialform}, which is required for the gauge
invariance of the WZ term \eqref{WZterm}, fixes the relation
\eqref{relation} between the RR-field $C_4$ and the
duality-covariant $A$-fields. Extending this to the higher-rank
$n$-forms the general WZ-term can be written in the universal
form \eqref{WZterm}. All $C$-fields are related in a
duality-covariant way to the $A$-fields, using the basic charges
${\tilde q}_\alpha, q_\alpha$. The world-volume 2-form curvature  ${\cal F}_2$ is given by
  \begin{equation}
  {\cal F}_2 = dV_1+B_2=q_\alpha {\cal F}_2^\alpha = q_\alpha\left( d V_1^\alpha + A_2^\alpha\right)\,.
  \end{equation}
Here we have introduced, together with the doublet of 2-form
potentials $A_2^\alpha$, a doublet of worldvolume vectors
$V_1^\alpha$. In this way the charges of all Type IIB D$p$-branes
with $p\ge 3$ can be expressed as products of the basic 1-brane
charges ${\tilde q}_\alpha, q_\alpha$, corresponding to D$1$-branes
and F$1$-strings, respectively \cite{Bergshoeff:2006gs}. The same
analysis shows that
   \begin{itemize}
   \item in the case of the 10-forms, which belong to a reducible
   representation of $\text{SL}(2,\mathbb{R})$, it is the quadruplet,
   that is the highest-dimensional representation, that contains
   the RR field $C_{10}$. One cannot write down a gauge-invariant WZ term for the
   doublet;

   \item for both the triplet of 8-forms and the quadruplet of 10-forms, not
   all $n$-forms  inside the representation can be reached by a duality rotation
   of the RR-field. More specifically, there is one 8-form out of the triplet of 8-forms and
   there are two 10-forms out of the quadruplet of 10-forms that cannot be viewed as
   a duality transformation of the RR $n$-form. For these potentials one
   cannot write down a gauge-invariant WZ term.
  \end{itemize}

It is well-known how the  $\text{SL}(2,\mathbb{R})$ symmetry of IIB
supergravity gets generalised to other symmetry groups for the
$D<10$ maximal supergravities \cite{Julia1,Julia2}. These symmetries
have been identified as the U-dualities, containing S- and
T-dualities, of superstring theory \cite{Hull:1994ys}. They have
also been discussed in the context of worldvolume actions of
extended objects in supergravity backgrounds \cite{Duff:1990hn}. The
different duality groups for $3\le D\le 11$ are given in Table 1.
The generic global symmetry group in $D$ dimensions ($D \leq 9$) can
be denoted (generalising what one gets in 3,4 and 5 dimensions) by
$\text{E}_{11-D}$, where $11-D$ is the rank of the group.

\begin{table}
\begin{center}
\begin{tabular}{||l|l||}
\hline
dimension $D$&duality group $G$\\
\hline
11&1\\
10A&$\mathbb{R}^+$\\
10B&$\text{SL}(2)$\\
9&$\text{GL}(2)$\\
8&SL(3)$\times$SL(2)\\
7&SL(5)\\
6&SO(5,5)\\
5&$\text{E}_6$\\
4&$\text{E}_7$\\
3&$\text{E}_8$\\
\hline
\end{tabular}
\end{center}
 \caption{\sl   The U-duality groups for all maximal supergravities in
dimensions $3\le D\le 11$. The group is always over the real numbers
and of split real form. In $D=10$ we distinguish between IIA and IIB
supergravity.} \end{table}

In each dimension the corresponding maximal supergravity theory
contains a number of $n$-forms that transform in given
representations of the U-duality group. These representations
naturally follow by making a level decomposition of the very
extended  Kac-Moody algebra $\text{E}_{11}$
\cite{West:2001as,Schnakenburg:2001ya,Kleinschmidt:2003mf,Riccioni:2007au,Bergshoeff:2007qi}.
Following the notation of \cite{Riccioni:2009xr}, for each $n$-form
we denote these representations with a lower $M_n$ index. The forms
are thus denoted by
  \begin{equation}\label{hierarchy}
  A_{1,M_1}\,,\hskip .5truecm A_{2,M_2}\,,\hskip .5truecm A_{3,M_3}\,,\hskip
  .5truecm \cdots \hskip .5truecm\,,A_{D-1,M_{D-1}}\,,\hskip .5truecm
  A_{D,M_{D}}\, .
  \end{equation}
All fields decompose into representations of the T-duality group
$\text{SO}(10-D,10-D)$, which is a subgroup of the U-duality group
$\text{E}_{11-D}$. It is convenient to make this decomposition since
all RR fields transform as irreducible representations of the
T-duality group. A convenient way to see whether a given $n$-form
potential is a RR field is by calculating the corresponding brane
tension. For a RR field this tension should scale as $1/g_s$ in the
string frame, with $g_s$ the string coupling constant. It turns out
that in each dimensions $D$ the RR fields of odd rank transform in
the spinor representation of $\text{SO}(10-D,10-D)$ and we denote
them with $C_{2n-1 , a}$, with $a$ an $\text{SO}(10-D,10-D)$ spinor
index, while the RR fields of even rank transform in the conjugate
representation and we denote them with $C_{2n , \dot{a}}$ \cite{Obers:1998fb}. Besides
the RR-fields we need to consider the Fundamental 2-forms and
1-forms that couple to the Fundamental string and Fundamental
0-branes, i.e.~wrapped Fundamental strings, respectively. These
Fundamental fields have corresponding brane tensions that are
independent of the string coupling constant, again in the string
frame. In each dimension the Fundamental 1-forms transform in the
vector representation of $\text{SO}(10-D,10-D)$ and we denote them
with $B_{1,A}$, while the Fundamental 2-form is a T-duality singlet,
$B_2$.

As we will discuss in detail in section 4, the equivalent of the
charges $\tilde{q}_\a,q_\alpha$ introduced in Type IIB string theory
will be charges $\tilde{q}^{M_1}_{a}, q^{M_1}_A$,
 such that
  \begin{equation}\label{basic}
  C_{1, a} = \tilde{q}^{M_1}_{ a} A_{1, M_1 }\,,\hskip 1truecm B_{1,A} = q_A^MA_{1, M_1}
  \end{equation}
define the RR and Fundamental 1-forms in a U-duality covariant way.
We will see in section 4 how all the charges of the
higher-dimensional D$p$-branes, with $p\ge 1$, can be expressed as
products of these basic 0-brane charges. Unlike in Type IIB string
theory, all charges can be expressed in terms of 0-brane charges
only, no 1-brane charges are involved. This has to do with the fact
that, for  $D<10$, the basic gauge symmetries generating the whole
gauge algebra are always the ones corresponding to the 1-form
potentials only.

The aim of this paper is to construct gauge-invariant and
duality-covariant expressions for general WZ terms using the
ingredients introduced above. These WZ terms will describe the
coupling of general D$p$-branes in dimensions $3\le D\le 10$ to the
target space supergravity fields in a duality-covariant way. For
$D=10$ there have been attempts to construct such WZ terms. For
instance, an $\text{SL}(2,\mathbb{R})$-invariant formulation of
1-branes has been given \cite{Townsend:1997kr,Cederwall:1997ts}.
This formulation made use of the fact that in two spacetime
dimensions the Born-Infeld vector is equivalent to an integration
constant describing the tension of a string. Similarly, the case of
3-branes has been discussed \cite{Cederwall:1997ab}. In this case
one makes use of the fact that in 4 spacetime dimensions the
electric-magnetic dual of a Born-Infeld vector is again a vector.
Such special properties do not occur for the branes with $p>3$ and
indeed constructing an $\text{SL}(2,\mathbb{R})$-invariant
formulation of 5-branes turns out to be problematic
\cite{Westerberg:1999fe}. For $D=10$ this gap was filled and
gauge-invariant and duality-covariant expressions for all the
D-branes of IIB string theory were given
\cite{Bergshoeff:2006gs}. In this paper we extend this work to
$D<10$ dimensions.

A basic difference with $D=10$ dimensions is that, to write down a
WZ-term in $D<10$ dimensions, we need to introduce, together with
the standard worldvolume scalars describing the position of the
brane in $D$ dimensions, not only a worldvolume vector $V_1$ but
also additional worldvolume scalars. This is due to the fact that
the Fundamental string can wrap around each of the $10-D$
compactified dimensions. Therefore, in $D<10$ not only strings but
also a number of particles can couple to the D-brane. The coupling
of these particles are described by the extra worldvolume scalars.
We find $2(10-D)$ of such scalars transforming as a vector $V_{0,A}$
under the T-duality group $\text{SO}(10-D,10-D)$, with corresponding
curvatures ${\cal F}_{1,A}$. In this paper we will show that, using
these curvatures, the general WZ term can be written in the
following elegant form:
  \begin{equation}\label{WZtermD<10}
  {\cal L}_{\text{WZ}}(\text{D} \le \text{10}) = e^{{\cal F}_2}e^{{\cal F}_{1,A}\Gamma^A}C\,,
  \end{equation}
where $\Gamma^A$ are the $\text{SO}(10-D,10-D)$ gamma matrices and
$C$ is a formal sum of all RR-potentials. Note that ${\cal
F}_{1,A}=0$ in $D=10$ since there is no T-duality in that dimension.
Therefore, the expression \eqref{WZtermD<10} reduces to the usual
expression \eqref{WZterm} for $D=10$.

To show that \eqref{WZtermD<10} is the correct gauge-invariant and
duality-covariant WZ term for any D$p$-brane in $D\le10$ dimensions,
our strategy will be as follows. First, for the convenience of the
reader, we will shortly review in Section 2 the WZ terms for the
D-branes of $D=10$ Type IIA and Type IIB string theory. Next,  we
will  present in Section 3 the general gauge algebra of maximal
supergravity in any dimension $3\le D\le 10$.  The structure of this
gauge algebra follows from the underlying $\text{E}_{11}$ algebra.
We will first give a general analysis and, next, work out the
formulae for each specific dimension. In Section 4 we will, starting
from the duality-covariant $A$-basis presented in  Section 3, derive
the expressions for the RR potentials $C$ that couple to the
D$p$-branes using the basic charge vectors \eqref{basic}. In
particular, we will give expressions for the charge vectors of the
higher-dimensional branes in terms of products of the basic charges
\eqref{basic}.  Like in Section 3, we will first give the general
analysis and then give explicit expressions for different
dimensions. Next, in Section 5 we will derive the main result of
this paper, i.e.~the gauge-invariant and duality-covariant WZ term
\eqref{WZtermD<10}. Finally, in Section 6 we will present our
conclusions and indicate a few natural extensions of this work.

\section{D-branes in Ten Dimensions}

For the convenience of the reader we shortly summarise in this
Section what is known about the WZ terms corresponding to the
D-branes of Type IIA and Type IIB superstring theory. We first
discuss the IIB case \cite{Bergshoeff:2006gs}.

\subsection{Type IIB}

Our starting point is the set of  $\text{SL}(2,\mathbb{R})$-covariant IIB $n$-form
potentials given in \eqref{A-fields}.
Note that the 4-form field $A_4$ in \eqref{A-fields} is self-dual
and that $A_2$ and $A_6$ are each other's dual. The triplet of
8-forms is dual to the  2 scalars (axion and dilaton) of IIB
supergravity.
The counting works (2 scalars are dual to 3 8-forms) since the
8-forms satisfy  a single constraint
\cite{Cremmer:1998px,Dall'Agata:1998va,Bergshoeff:2005ac}.
We next introduce the basic charges
$q_\alpha, {\tilde q}_\alpha$ as in \eqref{basicIIBcharges}, where
$q_\alpha$ and $\tilde{q}_\alpha$ are such that $\epsilon^{\alpha
\beta} \tilde{q}_{\alpha} q_\beta \neq 0$. This fixes our choice of
the RR potential $C_2$ and NS-NS potential $B_2$. We are using  the
normalisations of \cite{Bergshoeff:2005ac}, and convert these
results in form language in the usual way:
 \begin{equation}
  A_n = \frac{1}{n!} A_{\m_1 ...\m_n} dx^{\m_1} \wedge dx^{\m_2}
  \wedge ...\wedge d x^{\m_n} \quad . \label{defofform}
  \end{equation}

Our first task is to select among all $n$-form fields of IIB
supergravity the RR-potentials who have the defining property that
they do not transform under the NS-NS gauge transformations or,
equivalently, that they couple to D-branes whose tensions scale as
$1/g_s$. To see how this works, consider the gauge transformation of
the 4-form potential $A_4$
 \begin{equation}
 \d A_4 = d \Lambda_3 - \frac{i}{16} \e_{\a\b} \Lambda^\a_1 F_3^\b
 \quad ,
 \end{equation}
which transforms both under RR and NS-NS gauge transformations, with
parameters ${\tilde q}_\alpha\Lambda^\alpha$ and
$q_\alpha\Lambda^\alpha$, respectively. One may verify that there is
a unique combination of $A_4$ and $A_2 A_2$ terms that is invariant
under the NS-NS gauge transformations. This combination defines the
RR 4-form potential $C_4$:
\begin{equation}
 C_4 = \tilde{q}_\a q_\b \bigl( - i \e^{\a\b} A_4 - \tfrac{1}{16} A_2^\a
 A_2^\b \bigr) \quad .
 \end{equation}
Applying the same procedure to the higher-form potentials we obtain
the following expressions for the RR-potentials in terms of the
$\text{SL}(2,\mathbb{R})$-covariant $A$-fields:
  \begin{eqnarray}
  & & C_6 = \tilde{q}_\a q_\b q_\g \bigl( - i \e^{\a\b} A_6^\g -
  \tfrac{4}{3}i \e^{\a\b} A_4 A_2^\g - \tfrac{1}{12} A_2^\a A_2^\b
  A_2^\g \bigr)\,, \nonumber \\[.2truecm]
  & & C_8 = \tilde{q}_\a q_\b q_\g q_\d \bigl( -i \e^{\a\b}
  A_8^{\g\d} - \tfrac{i}{16} \e^{\a\b} A_6^\g A_2^\d - \tfrac{i}{12}
  \e^{\a\b} A_4 A_2^\g A_2^\d - \tfrac{1}{192} A_2^\a A_2^\b A_2^\g
  A_2^\d \bigr)\,, \nonumber \\[.2truecm]
  & & C_{10} = \tilde{q}_\a q_\b q_\g q_\d q_\e \bigl( -i \e^{\a\b}
  A_{10}^{\g\d\e} + \tfrac{i}{15} \e^{\a\b}A_8^{\g\d} A_2^\e +
  \tfrac{i}{240}\e^{\a\b}
  A_6^\g A_2^\d A_2^\e + \tfrac{i}{180}\e^{\a\b} A_4 A_2^\g A_2^\d A_2^\e \nonumber \\
  & & \quad \quad +
  \tfrac{1}{2880} A_2^\a A_2^\b A_2^\g A_2^\d A_2^\e \bigr)\,.
  \label{summaryofIIB}
  \end{eqnarray}
These RR $n$-form potentials occur as the representations ${\bf
1}_{n/2-2}$ in the decomposition of the $\text{SL}(2,\mathbb{R})
$-covariant $A$-fields under

\begin{equation}
\text{SL}(2,\mathbb{R}) \supset \mathbb{R}^+\,,
\end{equation}
where the sub-index $n/2 -2$ indicates the $\mathbb{R}^+$-- charge $w$
of the representation. The complete decomposition is given in Table
\ref{D=10IIB}. Each field can be associated with a $p$-brane
($p=n-1$) if it exists, whose brane tension in string frame scales
as\,\footnote{This general formula only applies if the $n$-form
transforms under supersymmetry to the gravitino with a non-zero
coefficient that only depends on the dilaton and not the axion.}
  \begin{equation}\label{generalformulaIIB}
  g_{s}^\alpha\,,\hskip 2truecm \alpha =
  \tfrac{1}{2}\left(-\tfrac{n}{2}+w\right)\,.
  \end{equation}
We distinguish between the following objects:
\begin{eqnarray*}
\alpha=-1&:&\ \ \ \text{D-brane}\,,\\
\alpha=0&:&\ \ \ \text{Fundamental Object}\,,\\
\alpha=-2&:&\ \ \ \text{Solitonic Object}\,,\\
\alpha<-2&:&\ \ \ \text{Rest}\,,
\end{eqnarray*}
which in the Table are given in the columns RR, F, S and Rest,
respectively.

 \begin{table}
\begin{center}
\begin{tabular}{|c|c||c|c|c|c|}
\hline \rule[-1mm]{0mm}{6mm} field & U repr & RR &  F & S& Rest\\
\hline \hline \rule[-1mm]{0mm}{6mm} 2-form & ${\bf 2}$ &
${\bf 1}_{-1}$ & ${\bf 1}_1$ && \\
\hline \rule[-1mm]{0mm}{6mm} 4-form & ${\bf 1}$ & ${\bf 1}_{0}$ &
 & &\\
\hline \rule[-1mm]{0mm}{6mm} 6-form & ${\bf 2}$ & ${\bf 1}_{1}$ & & ${\bf 1}_{-1}$&\\
\hline \rule[-1mm]{0mm}{6mm} 8-form & ${\bf 3}$ & ${\bf 1}_2$ & & $``{\bf 1}_{0}$'' & ${\bf 1}_{-2}$\\
\hline \rule[-1mm]{0mm}{6mm} 10-form & ${\bf 4}$ & ${\bf 1}_3$ & & $``{\bf 1}_{1}$'' & $``{\bf 1}_{-1}$'' $+ {\bf 1}_{-3}$\\
\cline{2-6} \rule[-1mm]{0mm}{6mm}  & ${\bf 2}$ &  & & $``{\bf 1}_1$'' & $``{\bf 1}_{-1}$''
\\
\hline
\end{tabular}
\end{center}
\caption{\sl {The ten-dimensional IIB case: the 2nd column indicates
the $\text{SL}(2,\mathbb{R})$ representation of the $A$-fields. The
3rd, 4th and 5th column indicate the Ramond-Ramond, Fundamental and
Solitonic fields, respectively. The last column contains all fields
with different dilaton couplings. It is not clear whether the
8-forms and 10-forms  indicated by accolades couple to a brane (see
the text).} \label{D=10IIB}}
\end{table}

We find the following expression of the charge of a D$p$-brane in
terms of the D1-brane and F1-brane charges:
\begin{equation}\label{dependent}
{\tilde q}_{\alpha_1\cdots \alpha_{m-1}} = \left({\tilde q}_\alpha
q_\beta \epsilon^{\alpha\beta}\right)q_{\alpha_1}\cdots
q_{\alpha_{m-1}}\,,\hskip 1truecm p=2m+1\,.
\end{equation}
Indicating with $n(q)$ and $n({\tilde q})$ the number of $q$ and
${\tilde q}$ charges that occur in the expression \eqref{dependent}
we have the following relations:
  \begin{equation}
  \alpha = -n({\tilde q})\,,\hskip 2truecm w = -n({\tilde q}) + n(q)\,.
  \end{equation}
Note that there are D$p$-branes for each odd $p$ but that there is
only one Fundamental string F1. The ${\bf 1}_1$ 6-form is dual to
the ${\bf 1}_{-1}$ 2-form and represents the duality between the
D5-brane and D1-brane. The ${\bf 1}_{-1}$ 6-form is dual to the
${\bf 1}_1$ 2-form and represents the duality between the NS 5-brane
NS5B and the Fundamental F1 string.

The ${\bf 1}_2$ 8-form couples to the D7-brane with charge
  \begin{equation}
  {\bf 1}_2\,:\ \ \ \left({\tilde q}_\alpha q_\beta\epsilon^{\alpha\beta}\right)q_\gamma q_\delta\,.
  \end{equation}
The other two 8-forms in the same triplet have charges:

\begin{equation}
{\bf 1}_0\,:\ \ \ \left({\tilde q}_\alpha q_\beta\epsilon^{\alpha\beta}\right){\tilde q}_{(\gamma} q_{\delta)}\,,\hskip 2truecm
{\bf 1}_{-2}\,:\ \ \ \left({\tilde q}_\alpha q_\beta\epsilon^{\alpha\beta}\right){\tilde q}_\gamma {\tilde q}_\delta\,.
\end{equation}
We observe that the ${\bf 1}_2$ D7-brane and its S-dual, the ${\bf
1}_{-2}$ S7-brane, both have charges that are proportional to the
product of two uncontracted $\text{SL}(2,\mathbb{R})$ vectors
(either $q_\alpha$ or $\tilde{q}_\a$) that are the same whereas the
charge corresponding to the ${\bf 1}_0$ 8-form contains the product
of two {\sl different} charges. This means that under a U-duality
the D7-brane and S7-brane can be transformed into each other but
that one can never rotate one of these branes into a brane
corresponding to the ${\bf 1}_0$ 8-form. In other words, the
D7-brane and S7-brane belong to the same conjugacy class that forms
a (non-linear) doublet embedded into the triplet.  The fact that
they belong to the same conjugacy class can also be deduced from the
fact that, viewed as a $2\times 2$ matrix, both charges have zero
determinant, i.e.~det\,$\left [q_\alpha q_\beta\right] = $
det\,$\left [{\tilde q}_\alpha {\tilde q}_\beta\right]=0.$ It is not
clear whether the ${\bf 1}_0$ 8-form couples to a brane since under
supersymmetry it transforms to the gravitino with an axion-dependent
coefficient, thereby violating one of the assumptions that go into
the general formula \eqref{generalformulaIIB}. What is
clear is that for such an object one can not write down a gauge-invariant WZ term as
in \cite{Bergshoeff:2006gs} because its charge is proportional to the
product of two {\sl different} uncontracted 1-brane charge vectors.

We finally consider the 10-forms. We first consider the quadruplet of 10-forms.
The ${\bf 1}_3$ 10-form potential
couples to the D9-brane and has charge
\begin{equation}
{\bf 1}_3\,:\ \ \  \left({\tilde q}_\alpha q_\beta\epsilon^{\alpha\beta}\right)q_\gamma q_\delta q_\epsilon\,.
\end{equation}
The other three objects in the quadruplet have charges
  \begin{equation}
  {\bf 1}_1:\ \  \left({\tilde q}_\alpha q_\beta\epsilon^{\alpha\beta}\right)
  {\tilde q}_{(\gamma} q_{\delta}q_{\epsilon)}\,,\hskip .7truecm
  {\bf 1}_{-1}:\  \ \left({\tilde q}_\alpha q_\beta\epsilon^{\alpha\beta}\right){\tilde q}_{(\gamma}
  {\tilde q}_\delta q_{\epsilon)}\,, \hskip .7truecm
  {\bf 1}_{-3}:\ \ \left({\tilde q}_\alpha
  q_\beta\epsilon^{\alpha\beta}\right){\tilde q}_{ \gamma} {\tilde q}_\delta {\tilde q}_{\epsilon }\,.
  \end{equation}
Applying the same reasoning as in the case of the 7-branes, we
conclude that the D9-brane is in the same conjugacy class as its
S-dual, the ${\bf 1}_{-3}$ brane since these are the only two
objects whose charge is proportional to the product of three
$\text{SL}(2,\mathbb{R})$ vectors $q_\alpha$ that are the same.
Together, they form a non-linear doublet embedded into the
quadruplet. It is not clear whether the other, ${\bf 1}_1$ and ${\bf
1}_{-1}$,  10-forms couple to a 9-brane since they violate the
assumptions underlying \eqref{generalformulaIIB}. Anyway, for these
two quantities it is impossible to write a gauge-invariant WZ term as in
\cite{Bergshoeff:2006gs}.

We next consider the charges associated to the doublet of ten-forms. These are
  \begin{equation}
  {\bf 1}_1:\ \  \left({\tilde q}_\alpha
  q_\beta\epsilon^{\alpha\beta}\right) \left(
  {\tilde q}_{\gamma} q_{\delta} \e^{\g\d} \right) q_{\epsilon }\,,\hskip .7truecm
  {\bf 1}_{-1}:\  \ \left({\tilde q}_\alpha q_\beta\epsilon^{\alpha\beta}\right)\left( {\tilde q}_{\gamma}
  {q}_\delta  \e^{\g\d} \right) \tilde{q}_{\epsilon}\,.
  \end{equation}
It turns out that using these expressions it is impossible to write
down a corresponding gauge-invariant WZ term along the lines of
\cite{Bergshoeff:2006gs}.

\subsection{Type IIA}

Although IIA superstring theory has only an $\mathbb{R}^+$-- duality
symmetry the situation is similar to the IIB case. In the IIA case
the charges $\tilde{q}^{(n)}$ of the higher-dimensional D$p$-branes
can be written as products of the D0-brane charge $\tilde q$ and the
Fundamental F1-string charge $q$ as follows:
\begin{equation}\label{dependent2}
\tilde{q}^{(m)} = {\tilde q}\,q^{m/2-1/2}\,,\hskip 1truecm m=p+1\,.
\end{equation}
Note that, unlike in the IIB case, the basic charges correspond not
only to 1-branes but also to 0-branes. The difference with the IIB
case can be traced back to  the fact that the IIA and IIB gauge
algebras have different so-called fundamental symmetries. The
fundamental symmetries  are the basic gauge symmetries out of which
all other symmetries can be generated by taking multiple
commutators. It turns out that the fundamental symmetries of the IIB
gauge algebra are the ones corresponding to the doublet of 2-form
potentials, i.e.~the RR and NS-NS 2-form potentials, whereas those
of the IIA gauge algebra are given by the gauge symmetries
corresponding to the RR 1-form and NS-NS 2-form potentials.

 \begin{table}
\begin{center}
\begin{tabular}{|c||c|c|c|}
\hline \rule[-1mm]{0mm}{6mm} field  & RR &  F & S\\
\hline \hline \rule[-1mm]{0mm}{6mm} 1-form  &
${\bf 1}_{-1}$ &  & \\
\hline \rule[-1mm]{0mm}{6mm} 2-form &  &${\bf 1}_2$
 & \\
\hline \rule[-1mm]{0mm}{6mm} 3-form  & ${\bf 1}_{1}$ & & \\
\hline \rule[-1mm]{0mm}{6mm} 5-form & ${\bf 1}_3$ & &  \\
\hline \rule[-1mm]{0mm}{6mm} 6-form &  & & ${\bf 1}_{2}$  \\
\hline \rule[-1mm]{0mm}{6mm} 7-form  & ${\bf 1}_5$ & &   \\
\hline \rule[-1mm]{0mm}{6mm} 8-form  &  & & $``{\bf 1}_4$''  \\
\hline \rule[-1mm]{0mm}{6mm} 9-form  & ${\bf 1}_7$ & &   \\
\hline \rule[-1mm]{0mm}{6mm} 10-form  &  & &   $``{\bf 1}_{6}$'' \\
\cline{2-4} \rule[-1mm]{0mm}{6mm}   &  & &   ``${\bf 1}_{6}$''
\\
\hline
\end{tabular}
\end{center}
\caption{\sl {The ten-dimensional IIA case: the 2nd, 3rd and 4th column indicates
the $\mathbb{R}^+$-- representations of the Ramond-Ramond, Fundamental and Solitonic fields, respectively. It is not clear
whether the 10-forms indicated by accolades couple to a brane.}
\label{D=10IIA}}
\end{table}

We have collected the different $n$-form potentials of IIA
supergravity in Table \ref{D=10IIA}. The corresponding brane
tensions scale as\,,\footnote{This formula assumes that the $n$-form
transforms to a gravitino with a non-zero coefficient.}
  \begin{equation}\label{generalformulaIIA}
  g_{s}^\alpha\,,\hskip 2truecm \alpha = -\tfrac{1}{2}\left(n-w\right)=-n({\tilde q})\,,
  \end{equation}
where $w$ is the weight under $\mathbb{R}^+$, $n$ is the rank of the
form and $n({\tilde q})$ is the number of ${\tilde q}$ charges that
occurs in the expression of the $p$-brane charge, with $p=n-1$. In the second
column we have indicated all RR $n$-form potentials with $n$ odd.
The 3d column contains the Fundamental 2-form that couples to the
Fundamental string F1. The ${\bf 1}_2$ 6-form in the 4th column
couples to the solitonic 5-brane NS5A that is dual to the
Fundamental F1 string. The ${\bf 1}_4$ 8-form is the dual of the IIA
dilaton. It is not clear whether it couples to a brane since under
supersymmetry it does not transform to the gravitino, which was one
of the assumptions going into \eqref{generalformulaIIA}. Finally,
there are two 10-forms. For both of them it is not clear whether
they couple to a brane \cite{Bergshoeff:2010mv}.

\section{Gauge Algebra in Any Dimension}
In \cite{Riccioni:2009xr} the gauge algebra of all maximal ungauged
and gauged supergravities in any dimension was derived from
$\text{E}_{11}$. Under $\text{E}_{11}$ the fields transform as
  \begin{equation}
  \delta A = a + a \wedge A + a \wedge A \wedge A + ... \quad ,
  \end{equation}
where we denote with $A$ the form fields and with $a$ the
corresponding constant parameters. In the ungauged case, which is
the one on which we will focus from now on, these transformations
are promoted to gauge transformations via the
identification\footnote{The algebraic setup underlying the promotion
of the global $\text{E}_{11}$ symmetries to local ones was
constructed in \cite{Riccioni:2009hi}. The same construction applies
to the case of gauged supergravities, in which the gauge algebra
results from the identification $a \rightarrow d \Lambda + g
\Lambda$, where $g$ is the gauge coupling constant
\cite{Riccioni:2009xr}.}
  \begin{equation}
  a \rightarrow d \Lambda \quad ,
  \end{equation}
and the resulting gauge algebra can be schematically written as
  \begin{equation}
  \delta A = d \Lambda + d \Lambda \wedge A + d \Lambda \wedge A \wedge A + ... \quad .
  \label{E11gaugealgebra}
  \end{equation}
Although the $\text{E}_{11}$ algebra is  non abelian, one can  make,
in any dimension, suitable field redefinitions, as well as
field-dependent redefinitions of the gauge parameters, such that the
resulting gauge algebra is abelian, that means that the gauge
transformations are gauge invariant.\,\footnote{In the context of
supergravity, the existence of  an abelian basis for the gauge
algebra has been used in \cite{Cremmer:1998px}. More generally,
gauge algebras that do not necessarily have an abelian basis have
been considered in \cite{Bergshoeff:2009ph}.}
 This corresponds to
writing the gauge transformations as
  \begin{equation}
  \delta A = d \Lambda + \Lambda \wedge F \quad ,
  \end{equation}
where $F$'s denote the gauge-invariant field strengths. In this section we will
derive the gauge algebra in this basis.

In $D$ dimensions the global symmetry is $\text{E}_{11-D}$, and each
$n$-form carries a representation of $\text{E}_{11-D}$ that we
denote with a lower $M_n$ index \cite{Riccioni:2009xr}, see eq.~\eqref{hierarchy}.
Using this notation the field strength of the 1-form is
  \begin{equation}
  F_{2,M_1} = d A_{1,M_1}
  \end{equation}
and it is invariant under the gauge transformation
  \begin{equation}
  \delta A_{1,M_1} = d \Lambda_{0, M_1} \quad .
  \end{equation}

The field strength of the 2-form is
  \begin{equation}
  F_{3,M_2} = d A_{2,M_2 } + f^{M_1 N_1}{}_{M_2} A_{1,M_1} F_{2,N_1} \quad ,
  \label{2formfieldstrength}
  \end{equation}
where $f^{M_1 N_1}{}_{M_2}$ is an invariant tensor of
$\text{E}_{11-D}$ and an upstairs index $M_n$ denotes the conjugate
representation, which means that there is an invariant tensor
$\delta^{M_n}_{N_n}$. Gauge invariance of $F_{3,M_2}$ implies
  \begin{equation}
  \delta A_{2,M_2} = d \Lambda_{1,M_2} - f^{M_1 N_1}{}_{M_2}
  \Lambda_{0,M_1} F_{2,N_1} \quad . \label{twoformgaugetransf}
  \end{equation}
One also gets the Bianchi identities
  \begin{eqnarray}
  & & d F_{2,M_1} =0 \nonumber \\
  & & d F_{3,M_2} = f^{M_1 N_1}{}_{M_2} F_{2,M_1 } F_{2,N_1} \quad .
  \end{eqnarray}
The part of the invariant tensor $f^{M_1 N_1}{}_{M_2}$ which is
antisymmetric in $M_1$ and $ N_1$ can always be eliminated by means of a
field redefinition $A_{2 ,M_2} \rightarrow A_{2 ,M_2} - \tfrac{1}{2}
f^{M_1 N_1}{}_{M_2} A_{1 , M_1} A_{1,N_1}$. We can therefore assume
that $f^{M_1 N_1}{}_{M_2}$ is symmetric in $M_1$ and $ N_1$. This condition
naturally follows from the Jacobi identities of the $\text{E}_{11}$
algebra.

The field strength of the 3-form is
  \begin{equation}
  F_{4,M_3} = d A_{3,M_3} + f^{M_1 M_2}{}_{M_3} A_{1,M_1} F_{3,M_2} +
  f^{M_2 M_1}{}_{M_3} A_{2,M_2} F_{2,M_1} \quad .
  \end{equation}
One can always redefine suitably the fields in such a way that the
$\text{E}_{11-D}$ invariant tensors satisfy the constraints
  \begin{eqnarray}
  & & f^{M_2 M_1}{}_{M_3} = 2 f^{M_1 M_2}{}_{M_3}\quad , \nonumber \\
  & & f^{( M_1 | M_2 |}{}_{M_3} f^{N_1 P_1 )}{}_{M_2} = 0 \quad ,
  \label{firstffconstraint}
  \end{eqnarray}
which we therefore assume. These can also be seen as coming from
$\text{E}_{11}$ Jacobi identities. To summarise, the field strength
is
  \begin{equation}
  F_{4,M_3} = d A_{3,M_3} + f^{M_1 M_2}{}_{M_3} [ A_{1,M_1} F_{4,M_2} +
  2 A_{2,M_2} F_{2,M_1}] \quad ,
  \end{equation}
and the gauge transformation of the 3-form is
  \begin{equation}
  \delta  A_{3,M_3} = d \Lambda_{2,M_3} - f^{M_1 M_2}{}_{M_3} [
  \Lambda_{0,M_1 } F_{3,M_2} + 2 \Lambda_{1,M_2} F_{2,M_1} ] \quad .
  \label{gaugetransf3formA}
  \end{equation}
One also has the Bianchi identity
  \begin{equation}
  d F_{4,M_3} = 3 f^{M_1 M_2}{}_{M_3} F_{2,M_1 } F_{3,M_2}
  \quad .
  \end{equation}

For the 4-form one gets
  \begin{equation}
  F_{5, M_4} = d A_{4,M_4} + f^{M_1 M_3}{}_{M_4} [A_{1,M_1 } F_{4,M_3} + 3
  A_{3,M_3} F_{2,M_1} ] + f^{M_2 N_2}{}_{M_4} A_{2,M_2 } F_{3,N_2}
  \end{equation}
and
  \begin{equation}
  \delta A_{4,M_4 } = d \Lambda_{3,M_4} - f^{M_1 M_3}{}_{M_4} [
  \Lambda_{0,M_1} F_{4,M_3} + 3 \Lambda_{2,M_3} F_{2,M_1} ] - f^{M_2
  N_2}{}_{M_4} \Lambda_{1,M_2} F_{3,N_2} \quad ,
  \end{equation}
with the constraints
  \begin{equation}
  f^{M_2 N_2}{}_{M_4} f^{M_1 N_1}{}_{N_2} = 6 f^{(M_1 | M_3
  |}{}_{M_4 } f^{N_1 ) M_2}{}_{M_3} \label{constraintff2}
  \end{equation}
and the constraint that $f^{M_2 N_2}{}_{M_4}$ is antisymmetric in $M_2$ and  $N_2$. This
analysis can easily be extended to forms of higher rank.

For the convenience of the reader we write the explicit form of the various invariant tensors
in dimension $D$ from 9 to 4, for each dimension separately. This  reproduces the results of
\cite{Riccioni:2009xr} in the abelian basis. The notations we use here are
 taken from that paper. The results are summarised in Table
\ref{formsinanydimandf}.
 \begin{table}
\begin{center}
\begin{tabular}{|c||c|c|c|c||c|c|c|c|}
\hline \rule[-1mm]{0mm}{6mm} $D$  &  $A_{1,M_1}$ & $A_{2,M_2}$ &
$A_{3,M_3}$ &
$A_{4,M_4}$ & $f^{M_1 N_1}{}_{M_2}$ & $f^{M_1 M_2}{}_{M_3}$ & $f^{M_1 M_3}{}_{M_4}$ & $f^{M_2 N_2}{}_{M_4}$ \\
\hline \hline \rule[-1mm]{0mm}{6mm} 9   & $A_{1 ,\alpha} \ A_1$ &
$A_{2 , \alpha}$ & $A_3$ & $A_4$ & $\delta^\alpha_\beta$ &
$\epsilon^{\alpha \beta}$ & $-\tfrac{1}{3}$
& $\epsilon^{\alpha\beta}$ \\
\hline  \rule[-1mm]{0mm}{6mm}
8  & $A_{1, M\a}$ & $A_{2}^M$ & $A_{3,
\a}$ & $A_{4 ,M}$ & $\epsilon^{MNP}\epsilon^{\a\b} $ & $\delta^{M}_N
\delta^{\a}_\b $ & $\tfrac{1}{12} \delta^M_N \e^{\a\b}$ & $\e_{MNP}$\\
\hline \rule[-1mm]{0mm}{6mm}
7  & $A_{1, MN}$ & $A_{2}^M$& $A_{3,
M}$& $A_{4}^{MN}$ & $\epsilon^{MNPQR}$ & $\delta^{[M}_P
\delta^{N]}_Q$ & $-\tfrac{1}{3} \epsilon^{MNPQR} $ & $\delta^{[ P}_M
\delta^{Q
  ]}_N$\\
\hline \rule[-1mm]{0mm}{6mm}
6  & $A_{1 ,\dot{\a}}$ & $A_{2,M}$ &
$A_{3, \a}$ & $A_{4, MN}$ & $(C \Gamma_M )^{\dot\a \dot\b}$ &
$(\Gamma^M)_\a{}^{\dot\a}$ & $-\tfrac{1}{12} (C \Gamma_{MN})^{\dot\a
\a}$ & $\delta^{[ P}_M \delta^{Q
  ]}_N $\\
\hline \rule[-1mm]{0mm}{6mm}
5  & $A_{1, M}$ & $A_{2}^M$ & $A_{3,
\a}$ & $A_4^{MN}$ & $d^{MNP}$ & $D_{\a ,M}{}^N$ & $S^{\a M,NP}$ &
$\delta^{[ P}_M \delta^{Q
  ]}_N$ \\
\hline \rule[-1mm]{0mm}{6mm}
4 &  $A_{1,M}$& $A_{2,\a}$ &
$A_{3,A}$ & $A_{4,\a\b}$ &
$D_{\a}^{MN}$  & $S^{M\a}_A$ & $C^{MA}_{\a\b}$ & $\delta^{[\a}_\g \delta^{\b ]}_\d$ \\
\hline
\end{tabular}
\end{center}
\caption{\sl The $\text{E}_{11-D}$ invariant tensors associated to
all the fields up to the 4-form corresponding to  dimensions $D$ from 9 to 4.
The definitions of the invariant tensors and the relations between them are
taken from \cite{Riccioni:2009xr}.} \label{formsinanydimandf}
\end{table}

\subsection*{D=9}

In nine  dimensions the global symmetry is $\text{GL}(2,\mathbb{R})=\text{SL}(2,\mathbb{R})\times \mathbb{R}^+$
and the fields are
\begin{center}
\begin{tabular}{l l l}
1-form & ${\bf 2_0 \oplus 1_1 }$ & $A_{1\alpha}$, $A_1$ \\
2-form & ${\bf 2}_1$ & $A_{2 ,\alpha}$ \\
3-form & ${\bf 1}_1$ & $A_{3}$ \\
4-form & ${\bf 1}_2$ & $A_{4}$ \ , \\
\end{tabular}
\end{center}
where $\alpha =1,2 $ is an $\text{SL}(2,\mathbb{R})$ doublet index and the sub-index indicates the $\mathbb{R}^+$-- charge.
The only non-trivial invariant tensor is the epsilon symbol of
$\text{SL}(2,\mathbb{R})$, and from Table \ref{formsinanydimandf}
one can see that the conditions \eqref{firstffconstraint} and
\eqref{constraintff2} are satisfied.

\subsection*{D=8}
In eight dimensions the global symmetry is $SL(3,\mathbb{R})
\times SL(2,\mathbb{R})$ and the fields are ($M=1,2,3; \alpha=1,2$)
\begin{center}
\begin{tabular}{l l l}
1-form & ${\bf (\overline{3},2) }$ & $A_{1, M \alpha}$ \\
2-form & ${\bf (3,1)}$ & $A_{2}^{M}$ \\
3-form & ${\bf (1,2)}$ & $A_{3, \a}$ \\
4-form & ${\bf (\overline{3},1)}$ & $A_{4 , M}$  \ .\\
\end{tabular}
\end{center}
Given the invariant tensors of Table \ref{formsinanydimandf}, the
conditions \eqref{firstffconstraint} and \eqref{constraintff2} are
identically satisfied.

\subsection*{D=7}
In seven dimensions the symmetry is $SL(5, \mathbb{R})$, while the
representations of the fields up to the 5-form and the corresponding
notations are ($M=1,\dots,5$)
\begin{center}
\begin{tabular}{l l l}
1-form & ${\bf \overline{10}}$ & $A_{1, MN}$ \\
2-form & ${\bf 5}$ & $A_{2}^{M}$ \\
3-form & ${\bf \overline{5}}$ & $A_{3, M}$ \\
4-form & ${\bf 10}$ & $A_{4 }^{ MN}$ \ .\\
\end{tabular}
\end{center}
Given the invariant tensors of Table \ref{formsinanydimandf}, one may verify that the
conditions \eqref{firstffconstraint} and \eqref{constraintff2} are
satisfied.

\subsection*{D=6}
The global symmetry of the six-dimensional theory is $SO(5,5)$. The
representations of the fields up to the 5-form, and their
corresponding notations, are ($\alpha,\dot\alpha = 1,\dots,16;
M=1,\dots,10$)
\begin{center}
\begin{tabular}{l l l}
1-form & ${\bf 16}$ & $A_{1, \dot{\a}}$ \\
2-form & ${\bf 10}$ & $A_{2 ,M}$ \\
3-form & ${\bf \overline{16}}$ & $A_{3, {\a}}$ \\
4-form & ${\bf 45}$ & $A_{4 , MN}$ \ .\\
\end{tabular}
\end{center}
The invariant tensors are 10-dimensional Gamma matrices, and the
conditions \eqref{firstffconstraint} and \eqref{constraintff2} are
Fierz identities.

\subsection*{D=5}
In five dimensions the global symmetry is $\text{E}_6$. The
representations of the fields up to the 5-form, and their
corresponding notations, are ($M=1,\dots,27; \alpha=1,\dots,78$)
\begin{center}
\begin{tabular}{l l l}
1-form & ${\bf 27}$ & $A_{1, M}$ \\
2-form & ${\bf \overline{27}}$ & $A_{2}^M$ \\
3-form & ${\bf 78}$ & $A_{3, \a}$ \\
4-form & ${\bf 351}$ & $A_{4 }^{MN}$ \ .\\
\end{tabular}
\end{center}
It should be noted that we are considering the split, or real form
of $\text{E}_6$, therefore the representations are real. With ${\bf
\overline{27}}$ we simply mean the representation that is dual to
the ${\bf 27}$. This will be the case in all dimensions. The
conditions \eqref{firstffconstraint} and \eqref{constraintff2} are
satisfied by the invariant tensors in Table \ref{formsinanydimandf}.

\subsection*{D=4}
Finally, in four dimensions the global symmetry is $E_7$. The
representations of the forms up to the 5-form and the corresponding
notations are ($M=1, \dots,56; \alpha=1,\dots,133; A=1,\dots,912$)
\begin{center}
\begin{tabular}{l l l}
1-form & ${\bf 56}$ & $A_{1, M}$ \\
2-form & ${\bf 133}$ & $A_{2 , \a}$ \\
3-form & ${\bf 912}$ & $A_{3, A}$ \\
4-form & ${\bf 8645} + {\bf 133}$ & $A_{4 , \a\b}$ \ .\\
\end{tabular}
\end{center}
The index $A$ with which we denote the ${\bf 912}$ representation should not be
confused with the $A$ vector index of the T-duality group in any
dimension. The reducible representation to which the 4-forms belong
is the antisymmetric product of two adjoint (i.e.~${\bf 133}$)
indices. Therefore the $\a\b$ indices of the 4-form are meant to be
antisymmetrised. Note that in all dimensions the representation of the 4-form
is the antisymmetrised product of two 2-form representations.

Given the invariant tensors listed in Table \ref{formsinanydimandf},
eq. \eqref{firstffconstraint} becomes
  \begin{equation}
  S^{\a (M}_A D_\a^{NP )} =0 \quad ,
  \end{equation}
while eq. \eqref{constraintff2} is
 \begin{equation}
 \delta^\a_{[\g} D^{MN}_{\d ]} = 6 C^{(M | A|}_{\g\d} S^{N) \a}_A
 \quad .
 \end{equation}
These two equations are both satisfied, see  \cite{Riccioni:2009xr}.

\section{From the $A$-fields to the $C$-fields}
In this section we wish to generalise the IIB construction  of
\cite{Bergshoeff:2006gs}, thus determining in any dimension the RR
$C$-fields starting from the duality-covariant $A$-basis of the
previous section. Schematically, we expect the $C$ fields to
transform under the RR gauge transformations as
  \begin{equation}
  \delta C \sim d \lambda +  H \wedge {\lambda} \quad ,
  \label{defofCbasis}
  \end{equation}
where $\lambda$ are the gauge parameters of the RR fields and $H$
are the field strengths of the Fundamental fields, that is the
fields corresponding to the branes whose tension does not depend on
the dilaton in the string frame. In ten dimensions, $H$ is a 3-form.
By dimensional reduction, we expect $H$ in $D<10$ dimensions to be
either a two-form or a three-form. That is why we have not yet
indicated the rank of the forms in \eqref{defofCbasis}. The precise
form of this equation will be given below, see
eq~\eqref{RRgaugetransfcompact}.

The T-duality subgroup of the U-duality group $\text{E}_{11-D}$ in
$D$ dimensions is $\text{SO}(10-D, 10-D)$. It is known that the RR
fields transform in the spinorial representations of the T-duality
group \cite{Obers:1998fb}. This is consistent with the fact that,
upon reduction over a single dimension, each D$p$-brane gives rise
to two branes: a D$p$-brane (dimensional reduction in the transverse
direction) and a D$(p-1)$-brane (dimensional reduction in the
worldvolume direction). Therefore, the total number of D-branes
doubles when going one down in the  dimension $D$. This is precisely
what happens with the dimension $d_{\text{spinor}}$ of a spinor
representation of the T-duality group $\text{SO}(10-D,10-D)$ which
is given by $d_{\text{spinor}} = 2^{9-D}$. It turns out that, more
precisely, the forms of odd rank transform as spinors of a given
chirality, while the forms of even rank transform as spinors of the
opposite chirality. As we will see in this section, this is
completely general and applies to RR $(D-1)$-- and $D$-- forms as well.

We now decompose in any dimension each duality-covariant form,
belonging to a given $\text{E}_{11-D}$ representation denoted by
$M_n$, in terms of representations of $\text{SO}(10-D, 10-D)$. More
precisely, we have
  \begin{equation}
  \text{E}_{11-D} \supset \text{SO}(10-D, 10-D) \times {\mathbb R}^+
  \label{UdualityTduality}
  \end{equation}
in all cases $D<10$. In $D=4$ and $D=3$ dimensions there are extra symmetry enhancements, such that in $D=4$  the
decomposition is
  \begin{equation}
  \text{E}_{7} \supset \text{SO}(6, 6) \times \text{SL}(2,\mathbb{R})
   \label{UdualityTdualityfourdim}
  \quad ,
  \end{equation}
and in $D=3$  one has
  \begin{equation}
  \text{E}_{8} \supset \text{SO}(8, 8)
   \label{UdualityTdualitythreedim}
  \quad .
  \end{equation}
 From the point of view of this paper these symmetry enhancements are not practical since they combine the RR--fields
 together with $n$--forms that couple to other kind of branes in one multiplet. We will therfore
 consider the further decompositions
\begin{equation}
\text{SL}(2,\mathbb{R}) \supset \mathbb{R}^+\label{furtherD=4}
\end{equation}
for $D=4$ and
\begin{equation}
 \text{SO}(8, 8) \supset \text{SO}(7, 7) \times \mathbb{R}^+\label{furtherD=3}
 \end{equation}
 for $D=3$.  In the
second part of this section we will discuss the above  decompositions for
each dimension separately. Here,  we first anticipate the main
outcome of this analysis.

Using the fact that  eq.~\eqref{defofCbasis}
relates a RR $n$-form to the gauge parameter of a RR $(n-2)$-form,
it follows that considering the decomposition
\eqref{UdualityTduality}, the ${\mathbb R}^+$-- charges
satisfy the relations
  \begin{equation}
  w_{\rm RR} ( n ) = w_{\rm RR} (n-1) + w_{\rm F} (1) \quad ,
  \qquad  w_{\rm RR} ( n ) = w_{\rm RR} (n-2) + w_{\rm F} (2) \quad
  , \label{qRR=qRRqNS}
  \end{equation}
where we denote with $  w_{\rm RR} ( n )$ the charge of the RR
$n$-form and with $w_{\rm F} (1)$ and $w_{\rm F} (2)$ the charge of
the Fundamental  1-forms and 2-forms. In Tables
\ref{qD=9}-\ref{qD=4} the decomposition of all the forms with
respect to T-duality is performed in any dimension $4\le D \le 9$.
From an analysis of the ${\mathbb R}^+$-- charges in the tables, it is
straightforward to see that the condition \eqref{qRR=qRRqNS} on the
charges has only one solution, and, moreover, it is the same
solution in any dimension. That solution is:
  \begin{itemize}
  \item All RR-forms, including  the $(D-1)$-forms and the $D$-forms,
  belong to the spinor representations of the T-duality group. The
  RR-forms of odd rank transform as spinors of a given chirality,
  and the RR-forms of even rank transform as spinors of the opposite
  chirality.
  \item In the cases in which the representation of the U-duality
  group is reducible, as is always the case for $D$-forms,
  and also for $(D-1)$-forms in dimension higher than six and for $(D-2)$-forms in dimension
  higher than seven, the RR-forms are always inside the highest-dimensional representation.
  \item The Fundamental 1-forms always belong to the vector representation of
  the T-duality group. We denote these fields by $B_{1 ,A}$, with
  $A$ a vector index of $\text{SO}(10-D, 10-D)$. Half of these represent the wrapped Fundamental strings,
  while the other half correspond to
  reduced pp-waves which, in $D=10$, are T-dual to the Fundamental string.
  \item The Fundamental 2-form always transforms as a singlet under
  T-duality. This is not surprising because this is the form
  associated to the Fundamental string.
  \end{itemize}
In the four-dimensional case, in which the decomposition of
eq.~\eqref{UdualityTdualityfourdim} occurs, the situation is more
subtle due to the further decomposition \eqref{furtherD=4} but the final result is exactly the same, as we will show in
the last subsection of this section.

The general result is summarised in Table
\ref{tablewithRRfields}.
  \begin{table}
\begin{center}
\begin{tabular}{|c|c|c|}
\hline \rule[-1mm]{0mm}{6mm} field & RR &  F \\
\hline \hline \rule[-1mm]{0mm}{6mm} $A_{1, M_1}$ & $C_{1 , a }$  & $B_{1 , A}$\\
\hline \rule[-1mm]{0mm}{6mm} $A_{2, M_2}$ & $C_{2 ,\dot{a} }$  & $B_{2} $\\
\hline \rule[-1mm]{0mm}{6mm} $A_{2n-1, M_{2n-1}}$ & $C_{2n-1 , {a} }$ \\
\cline{1-2} \rule[-1mm]{0mm}{6mm} $A_{2n , M_{ 2n}}$ & $C_{2n , \dot{a} }$ \\
\cline{1-2}
\end{tabular}
\end{center}
  \caption{\sl The RR and Fundamental fields in any dimension. In the last two lines $n$ is meant to be greater than 1.
\label{tablewithRRfields}}
\end{table}
A straightforward way of reaching this conclusion for the lower rank
forms is to consider explicitly the dimensional reduction of IIA and
IIB supergravity. The same dimensional reduction also gives the
right answer for the higher rank RR-forms that do not describe
physical degrees of freedom. It turns out that all  RR-forms in
$D<10$ dimensions arise from dimensional reduction of the IIA or IIB
RR-forms. This is not the case for the forms that we have collected
in the `Rest' columns of the different Tables which have dilaton
couplings different from the RR and Fundamental fields. Their
higher-dimensional origin resides in the mixed representations
predicted by $\text{E}_{11}$. These `Rest' fields are a minority in
$D=10$ dimensions but become a majority in lower dimensions. Besides
Solitonic objects they might describe other exotic objects in string
theory, with unconventional dilaton couplings, like $1/g_s^3,
1/g_s^4$, etc. The precise meaning of these objects, if they exist
et all, is not understood.

Given these general results, we now proceed with analysing how the
$C$-basis is defined in terms of the duality-covariant $A$-basis. As
we have already stressed and as Table \ref{tablewithRRfields} shows,
the RR-fields transform in the fermionic representations of
$\text{SO}(10-D, 10-D)$ and we denote them by
  \begin{equation}
  C_{2n-1 ,a} \qquad , \qquad \quad C_{2n , \dot{a}}
  \end{equation}
denoting with $a$ and $\dot{a}$  the $2^{9-D}$-dimensional spinor
representations of $\text{SO}(10-D, 10-D)$. It is useful to list the
conventions for the $\text{SO}(10-D,10-D)$ Gamma matrices that we
are using. In particular, we are using a Weyl basis, so that the
Gamma matrices have the form
  \begin{equation}
  \Gamma_A =  \left( \begin{array}{cc}
  0  &  (\Gamma_A)_a{}^{\dot{b}} \\
  (\Gamma_A)_{\dot{a}}{}^b & 0 \end{array} \right) \quad ,
  \end{equation}
where $a,\dot{a}=1, ...,2^{9-D}$. They satisfy the Clifford algebra
  \begin{equation}
  \{ \Gamma_A , \Gamma_B \} = 2 \eta_{AB}
  \end{equation}
where $\eta_{AB}$ is the Minkowski metric.

Because of ${\mathbb R}^+$-- charge conservation, we expect the NS-NS
field strengths of $B_{1, A}$ and $B_2$ to be
  \begin{eqnarray}
  & & H_{2 , A} = d B_{1, A} \nonumber \\
  & & H_3 = d B_2 + B_{1, A} H_{2, B} \eta^{AB} \quad ,
  \label{NSfieldstrengths}
  \end{eqnarray}
whose gauge invariance fixes the gauge transformations of the fields
to be
  \begin{eqnarray}
  & & \delta B_{1 ,A} = d \Sigma_{0, A} \nonumber \\
  & & \delta B_2 = d \Sigma_1 - \Sigma_{0 ,A} H_{2, B} \eta^{AB}
  \quad . \label{NSgaugetransfs}
  \end{eqnarray}
We write down the field strengths and gauge transformations of all
the RR fields in the compact form
  \begin{equation}
  G = d C + H_3 C + H_{2 ,A} \Gamma^A C
  \label{RRfieldstrengthcompact}
  \end{equation}
and
  \begin{equation}
  \delta C = d \lambda + H_3 \lambda - H_{2 ,A} \Gamma^A \lambda
  \quad . \label{RRgaugetransfcompact}
  \end{equation}
Here we denote with $C$ the sum of all the RR forms, in the spinor
representation of $\text{SO}(10-D,10-D)$, where each odd form in the
sum is projected on one chirality and each even form on the opposite
chirality, and the opposite projection occurs for the gauge
parameters $\lambda$. In components the two equations read
   \begin{eqnarray}
   & & G_{2n, a} = d C_{2n-1 ,a} + H_3 C_{2n-3 ,a} + H_{2 ,A}
   (\Gamma^A)_a{}^{\dot{b}} C_{2n-2 ,\dot{b}} \quad , \nonumber \\
   & & G_{2n+1 ,\dot{a}} = d C_{2n ,\dot{a} } + H_3 C_{2n-2
   ,\dot{a}} + H_{2 ,A} (\Gamma^A)_{\dot{a}}{}^b C_{2n-1 ,b } \quad ,
   \end{eqnarray}
and
  \begin{eqnarray}
  & & \delta C_{2n-1 , a} = d \lambda_{2n-2 ,a} + H_{3} \lambda_{2n-4 ,
  a } - H_{2 , A} (\Gamma^A)_a{}^{\dot{b}} \lambda_{2n-3 , \dot{b}}\quad ,
  \nonumber \\
  & & \delta C_{2n ,\dot{a}} = d \lambda_{2n-1 , \dot{a}} + H_3
  \lambda_{2n-3, \dot{a}} - H_{2 ,A} (\Gamma^A)_{\dot{a}}{}^b
  \lambda_{2n-2 , b} \quad .
  \end{eqnarray}

We now introduce the charges $\tilde{q}^{M_1}_a$ and $q^{M_1}_A$
that project the 1-forms on the RR and Fundamental 1-forms
respectively. That is
  \begin{equation}
  C_{1, a} = \tilde{q}^{M_1}_a A_{1,M_1} \quad , \qquad B_{1, A} =
  q^{M_1}_A A_{1,M_1} \quad .
  \end{equation}
Up to field redefinitions, the most general expression for the RR
and Fundamental 2-forms is
  \begin{eqnarray}
  & & B_2 = q^{M_2} A_{2, M_2} \nonumber \\
  & & C_{2, \dot{a}} = \tilde{q}^{M_2}_{\dot{a}} A_{2, M_2} + a
  (\Gamma^A)_{\dot{a}}{}^b \tilde{q}^{M_1}_b q^{N_1}_A A_{1, M_1}
  A_{1, N_1}
  \quad ,
  \end{eqnarray}
where the parameter $a$ will now be determined by consistency.
Indeed, varying both expressions according to eqs.
\eqref{twoformgaugetransf}, \eqref{NSgaugetransfs} and
\eqref{RRgaugetransfcompact}, and using the fact that $f^{M_1
N_1}{}_{M_2}$ is symmetric in $M_1$ and  $N_1$, one obtains $a=
\tfrac{1}{2}$. Furthermore, consistency implies the constraints
  \begin{equation}
  \tilde{q}^{M_2}_{\dot{a}} f^{M_1 N_1}{}_{M_2} =
  (\Gamma^A)_{\dot{a}}{}^b \tilde{q}^{( M_1}_b q^{N_1 )}_A
  \end{equation}
and
  \begin{equation}
  q^{M_2} f^{M_1 N_1}{}_{M_2} =q^{M_1}_A q^{N_1}_B \eta^{AB}
  \quad .
  \end{equation}
These relations can be inverted using the invariant tensor
$\tilde{f}_{M_1 N_1}{}^{M_2}$, where in general $\tilde{f}_{M_m
N_n}{}^{P_{m+n}}$ is such that
  \begin{equation}
  f^{M_m N_n}{}_{P_{m+n}} \tilde{f}_{M_m N_n}{}^{Q_{m+n}} =
  \delta_{P_{m+n}}^{Q_{m+n}} \quad ,
  \end{equation}
and thus they determine the charges of the 2-forms entirely in terms
of the charges of the 1-forms:
  \begin{eqnarray}
  & & \tilde{q}^{M_2}_{\dot{a}} =
  (\Gamma^A)_{\dot{a}}{}^b \tilde{q}^{ M_1}_b q^{N_1 }_A
  \tilde{f}_{M_1 N_1}{}^{M_2}\,, \nonumber \\
  & &
  q^{M_2} =q^{M_1}_A q^{N_1}_B \eta^{AB} \tilde{f}_{M_1 N_1}{}^{M_2}
  \quad . \label{expressionsforqtwos}
  \end{eqnarray}
To summarise, the expressions for the gauge fields and the gauge
parameters are
  \begin{eqnarray}
  & & B_2 =  q^{M_1}_A q^{N_1}_B \eta^{AB} \tilde{f}_{M_1 N_1}{}^{M_2} A_{2, M_2}\,, \nonumber \\
  & & C_{2, \dot{a}} = (\Gamma^A)_{\dot{a}}{}^b \tilde{q}^{M_1}_b
  q^{N_1}_A \bigl( \tilde{f}_{M_1 N_1}{}^{M_2} A_{2, M_2}  + \tfrac{1}{2} A_{1, M_1}
  A_{1, N_1} \bigr)\,,
  \end{eqnarray}
and
  \begin{eqnarray}
   & & \Sigma_1 = q^{M_2} \Lambda_{1, M_2}\,, \nonumber \\
   & & \lambda_{1, \dot{a}} = \tilde{q}^{M_2}_{\dot{a}} \Lambda_{1, M_2} +
   (\Gamma^A)_{\dot{a}}{}^b \tilde{q}^{[ M_1}_b q^{N_1 ]}_A \Lambda_{0,M_1}
   A_{1, N_1}
  \quad .
  \end{eqnarray}


We now move to the 3-forms. Before determining the actual expression
for the RR 3-form in terms of the $A$-fields, we first analyse the
corresponding charge. In principle, the charge $q^{M_3}_{a}$ can be
written as either
  \begin{equation}
  (\Gamma^A)_a{}^{\dot{b}} \tilde{q}^{M_2}_{\dot{b}} q^{M_1}_A
  \tilde{f}_{M_1 M_2}{}^{M_3}
  \end{equation}
or
  \begin{equation}
  \tilde{q}^{M_1}_a q^{M_2} \tilde{f}_{M_1 M_2}{}^{M_3} \quad .
  \end{equation}
Substituting the relations \eqref{expressionsforqtwos} and using the
condition
  \begin{equation}
\tilde{f}_{( M_1 | M_2 |}{}^{M_3} \tilde{f}_{N_1 P_1 )}{}^{M_2} = 0
\quad ,
  \end{equation}
which is the conjugate of the second of the constraints of
eq.~\eqref{firstffconstraint}, one can see that the two expressions
for $q^{M_3}_{a}$ are actually proportional, and therefore one can
write
    \begin{equation}
   \tilde{q}^{M_3}_a \propto \tilde{q}^{M_1}_{a} q^{N_1}_A q^{P_1}_B \eta^{AB}
   \tilde{f}_{N_1 P_1}{}^{M_2}
  \tilde{f}_{M_1 M_2}{}^{M_3} \quad .\label{q3q1q1q1}
  \end{equation}
We are now ready to determine the relation between the RR 3-forms
and the $A$-fields, which also determines the coefficient in
\eqref{q3q1q1q1}. The procedure is completely general: we know the
variation of the $C$ field, that is
eq.~\eqref{RRgaugetransfcompact}, and we known the transformation of
the 3-form $A$ field, that is eq.~\eqref{gaugetransf3formA}. We then
compare the two expressions solving for $C$ in terms of $A$. The
final result is
  \begin{eqnarray}
   & & C_{3, a} = \tilde{q}^{M_1}_b q^{N_1}_A q^{P_1}_B [
   -\tfrac{1}{2} \delta^b_a \eta^{AB} \tilde{f}_{N_1 P_1}{}^{M_2}
   \tilde{f}_{M_1 M_2}{}^{M_3} A_{ 3, M_3} - \tfrac{1}{3} ( \Gamma^A
   \Gamma^B )_a{}^b \tilde{f}_{M_1 P_1}{}^{M_2} A_{2, M_2} A_{1
   ,N_1} \nonumber \\
   & & \quad -\tfrac{2}{3} \delta^b_a \eta^{AB} \tilde{f}_{N_1
   P_1}{}^M_2 A_{1, M_1} A_{2 ,M_2} + \tfrac{1}{6}
   (\Gamma^{AB})_a{}^b A_{1,M_1} A_{1,N_1}A_{1,P_1} ] \quad .
   \end{eqnarray}

Proceeding this way one can determine the relation for all the RR $C$-fields
in terms of the duality-covariant $A$-fields. These results are
general and apply to any dimension. We give the expression for the
charge of any D-brane in terms of the charges $\tilde{q}^{M_1}_a$
and $q^{M_1}_A$. For instance, for the 4-form one gets
 \begin{equation}
 \tilde{q}^{M_4}_{\dot{a}} \propto (\Gamma^A)_{\dot{a}}{}^b \eta^{BC}
 \tilde{q}^{M_1}_b q^{N_1}_A q^{P_1}_B q^{Q_1}_C \tilde{f}_{P_1
 Q_1}{}^{M_2} \tilde{f}_{N_1 M_2}{}^{M_3} \tilde{f}_{M_1
 M_3}{}^{M_4} \quad .
 \end{equation}
This expression is unique, in the sense that there is no other
independent way of contracting one $\tilde{q}^{M_1}_a$ with three
$q^{M_1}_A$'s to get an object with the right indices. The reason
is, like in the discussion of the 3-forms above, the constraints
that the $\tilde{f}$ generalised structure constants satisfy for
consistency of the gauge algebra. In this case the relevant
constraint is the conjugate of eq.~\eqref{constraintff2}. One can
show that this is true in all cases, and the general expression for
the charge is
  \begin{equation}
  \tilde{q}^{M_{2n+1}}_a \propto \tilde{q}^{M_1}_a q^{N^{(1)}_1}_{A^{(1)}}
  q^{P^{(1)}_1}_{B^{(1)}}\eta^{A^{(1)}B^{(1)}} ...
  q^{N^{(n)}_1}_{A^{(n)}}
  q^{P^{(n)}_1}_{B^{(n)}}\eta^{A^{(n)}B^{(n)}} \tilde{f}_{M_1
  N^{(1)}_1}{}^{M_2} ... \tilde{f}_{ P^{(n)}_1 M_{2n}}{}^{M_{2n+1}}
  \end{equation}
for odd forms and
  \begin{equation}
  \tilde{q}^{M_{2n+2}}_{\dot{a}} \propto (\Gamma^A)_{\dot{a}}{}^b \tilde{q}^{M_1}_b
  q ^{N_1}_A q^{N^{(1)}_1}_{A^{(1)}}
  q^{P^{(1)}_1}_{B^{(1)}}\eta^{A^{(1)}B^{(1)}} ...
  q^{N^{(n)}_1}_{A^{(n)}}
  q^{P^{(n)}_1}_{B^{(n)}}\eta^{A^{(n)}B^{(n)}} \tilde{f}_{M_1
  N_1}{}^{M_2} ...
  \tilde{f}_{P^{(1)}_1 M_{2n+1}}{}^{M_{2n+2}}
  \end{equation}
for even forms.

Clearly one can in general construct charges with $n(\tilde{q}) \neq
1$, which do not correspond to D-branes. It turns that that there is
a simple formula for the dilaton scaling of the tension of the brane
to which the $n$-forms couple. It is given in terms of the number
 $n({\tilde q})$ of basic
${\tilde q}$ charges that one uses in the above expressions. All
tensions scale as
\begin{equation}
g_s^\alpha\,,\hskip 2truecm \alpha = -n({\tilde q})\,.\label{original}
\end{equation}
This formula assumes that the corresponding $n$-form transforms
under supersymmetry to the gravitino with a non-zero coefficient
that only depends on the dilaton scalar.

The structure for the charges we found above simplifies if we
consider only those charges that are generated by 2-form basic
charges. This corresponds to considering the gauge algebra of the
even-form fields whose gauge transformations are generated by the
2-form gauge transformation, that is considering only the structure
constants $f^{M_m N_n}{}_{P_{m+n}}$ where $m$ and $n$ are even. The
2-brane charges are in fact given in terms of the 1-form charges in
\eqref{expressionsforqtwos}, but now we consider them as basic
charges and we indicate them schematically with $Q$ in order to
distinguish them from the original 1-form charges $q$. We will make
use of this observation when we discuss the issue of conjugacy
classes for the $D=8$ case below. The simplification is that when
one projects on this sector, in the original expression for the
higher-form charges in terms of the 1-form basic charges $q$ all
1-form basic charges pair up to 2-form basic charges $Q \sim qq$.
More specifically, assuming that the 2-form sector only contains RR
and Fundamental 2-forms, which is true for $D>6$,\footnote{For
$D\leq 6$ one should include the charge of the solitonic 1-brane as
well, but this does not affect the analysis of the RR sector.} the
basic charges are given by
\begin{equation}
{\tilde Q}_{\dot a}^{M_2}\,,\hskip 2truecm Q^{M_2}\,.
\end{equation}
These basic charges define the RR and Fundamental 2-forms as
\begin{equation}
C_{2,\dot a} = {\tilde Q}_{\dot a}^{M_2} A_{2,M_2}\,,\hskip 2truecm B_2 = Q^{M_2}A_{2,M_2}\,.
\end{equation}
We next consider the  gauge algebra  restricted to this sector and,
for simplicity, we consider only the gauge transformations of the
2-forms and  4-forms:
  \begin{eqnarray}
  & & \d A_{M_2}   = d\Lambda_{M_2} \nonumber \\
  & & \d A_{M_4} = \d \Lambda_{M_4} - f^{M_2 N_2}{}_{M_4}
  \Lambda_{M_2} F_{N_2}
  \quad ,
  \end{eqnarray}
where  $f^{M_2 N_2}{}_{M_4}$ is antisymmetric in $M_2 N_2$. The
remarkable thing about this subsector is  that in all cases, that is
in any dimensions, the invariant tensor $f^{M_2 N_2}{}_{M_4}$ does
not put any constraint, that is the representation $M_4$ is in all
cases precisely the antisymmetric product of two $M_2$
representations \cite{Riccioni:2009xr}.  The consistency of the
gauge transformations requires the $C_{4 , \dot{a}}$ RR-field to be
expressed in terms of the U-duality-covariant $A$-fields as
  \begin{equation}
  C_{4 , \dot{a}} = \tilde{Q}^{M_2}_{\dot{a}} Q^{N_2} \bigl(
  \tilde{f}_{M_2 N_2}{}^{M_4} A_{4,M_4} - \frac{1}{2} A_{2,M_2 } A_{2, N_2}
  \bigr) \quad ,
  \end{equation}
where the charge $Q^{M_4}_{\dot{a}}$ is
  \begin{equation}
  \tilde{Q}^{M_4}_{\dot{a}} = \tilde{Q}^{M_2}_{\dot{a}} Q^{N_2} \tilde{f}_{M_2
  N_2}{}^{M_4} \quad .\label{Q4}
  \end{equation}
The  difference with respect to the full case is that no invariant
tensor of $\text{SO}(10-D, 10-D)$ takes part in the expression
\eqref{Q4}. This is because  the Fundamental 2-form charge is a
singlet. The only reason why this is consistent is that neither does
$\tilde{f}_{M_2 N_2}{}^{M_4}$ pose any constraint on the
representations. We therefore arrive at the conclusion that there is
a collaboration between the universal unconstrained structure of the
gauge algebra when restricted to the forms generated by the
even-form basic charges only,
 and the fact that there are fundamental strings in the
theory that are singlets under T-duality. In other words, the gauge
algebra knows about strings!

Below we show how things work out for each dimension $4 \le D \le 9$
separately, starting with the highest dimension.

\subsection*{D=9}
The U-duality symmetry of maximal supergravity in nine dimensions is
$\text{SL}(2, \mathbb{R})\times \mathbb{R}^+$.  One can consider the
nine-dimensional theory (as well as any lower dimensional one) as
coming from dimensional reduction of either the IIA or the IIB
theory. It is instructive to review how the branes in the IIA theory
can be seen from 11 dimensions, which is summarised as
   \begin{eqnarray}
   & & g_\m{}^\sharp \rightarrow {\rm D0} \qquad \quad \quad  A_{\m \n
   \sharp} \rightarrow {\rm F1} \nonumber \\
   & & A_{\m\n\r} \rightarrow {\rm D2}\quad \   \quad \quad   A_{\m_1 \dots \m_5 \sharp}
   \rightarrow {\rm D4} \nonumber \\
   & & A_{\m_1 \dots \m_6} \rightarrow {\rm NS5A} \quad .
   \label{IIAassignments}
   \end{eqnarray}
Here we denote with $\sharp$ the compact 11th coordinate, and the
6-form $A_6$ is the magnetic dual of the 3-form $A_3$ in eleven
dimensions.

Reducing to nine dimensions results in the fields collecting in
$\text{SL}(2, \mathbb{R})$  multiplets, as it is obvious from the
11-dimensional or IIB origin of the theory. We analyse the fields
from the IIA/11-dimensional viewpoint.\footnote{ From the IIB point
of view one obtains the same results, only the ten-dimensional
origin is different. What  is a reduced pp-wave from the IIA point
of view becomes a wrapped Fundamental string from the IIB point of
view. Nothing changes for RR fields.} Denoting with 9 the compact
10th coordinate, the 1-forms, and the corresponding 0-branes, are
  \begin{equation}
  ( g_\m{}^\sharp , g_\m{}^9 ) \rightarrow ({\rm D0 , F0)} \quad
  \quad A_{\m 9\sharp } \rightarrow {\rm F0} \quad ,
  \label{D=9D0F0F0}
  \end{equation}
while the 2-forms, and the corresponding 1-branes, are
  \begin{equation}
  ( A_{\m\n 9} , A_{\m\n \sharp} ) \rightarrow ( {\rm D1 ,F1 })
  \quad ,
  \label{D=9D1F1}
  \end{equation}
the 3-form (and 2-brane) is
  \begin{equation}
  A_{\m\n\r} \rightarrow {\rm D2}  \label{D=9D2}
  \end{equation}
and the 4-form (and 3-brane) is
  \begin{equation}
  A_{\m_1 \dots \m_4 9 \sharp} \rightarrow {\rm D3} \quad . \label{D=9D3}
  \end{equation}
The fact that we have called F0 the brane associated to the field
$g_\m{}^9$, which is a reduced pp-wave, is straightforward from considering
the T-dual IIB picture. For all the other cases the assignments are
straightforward from the reduction of eq. \eqref{IIAassignments}.
\begin{table}[!h]
\begin{center}
\begin{tabular}{|c|c||c|c|c|}
\hline \rule[-1mm]{0mm}{6mm} field & U repr & RR &  F & Rest\\
\hline \hline \rule[-1mm]{0mm}{6mm} 1-form & ${\bf 2}_0 $ &
$(-1,0)$ & $(1,0)$ &  \\
\cline{2-5} \rule[-1mm]{0mm}{6mm}  & ${\bf 1}_1$ &  & $(0,1)$ & \\
\hline \rule[-1mm]{0mm}{6mm} 2-form & ${\bf 2}_1$ & $(-1,1)$ &
$(1,1)$ & \\
\hline \rule[-1mm]{0mm}{6mm} 3-form & ${\bf 1}_1$ & $(0,1)$ & & \\
\hline \rule[-1mm]{0mm}{6mm} 4-form & ${\bf 1}_2 $
& $(0,2)$ & & \\
\hline \rule[-1mm]{0mm}{6mm} 5-form & ${\bf 2}_2$ & $(1,2)$ & & $(-1,2)$ \\
\hline \rule[-1mm]{0mm}{6mm} 6-form & ${\bf 2}_3$ & $(1,3)$ & &  $(-1,3)$ \\
\cline{2-5} \rule[-1mm]{0mm}{6mm}  & ${\bf 1}_2$ &  & & $(0,2)$\\
\hline \rule[-1mm]{0mm}{6mm} 7-form & ${\bf 3}_3$ & $(2,3)$ & & $(0,3) + (-2,3)$ \\
\cline{2-5} \rule[-1mm]{0mm}{6mm}  & ${\bf 1}_3$ & & & $(0,3)$\\
\hline \rule[-1mm]{0mm}{6mm} 8-form & ${\bf 3}_4$ & $(2,4)$ & &
$(0,4) + (-2,4)$ \\
\cline{2-5} \rule[-1mm]{0mm}{6mm}  & ${\bf 2}_3$ & & & $(1,3) + (-1,3)$\\
\hline \rule[-1mm]{0mm}{6mm} 9-form & ${\bf 4}_4$ & $(3,4)$ & &
$(1,4) + (-1,4)+ (-3,4)$ \\
\cline{2-5} \rule[-1mm]{0mm}{6mm}  & $2 \times{\bf 2}_4$ & & & $ 2\times [(1,4) + (-1,4)]$\\
 \hline
\end{tabular}
\end{center}
  \caption{\sl The decomposition of the $n$-form potentials of $D=9$ maximal
supergravity. The U-duality is $\text{SL}(2, \mathbb{R})\times
\mathbb{R}^+$. We denote with $(w_1, w_2 )$ the weights associated
to $\mathbb{R}^+\times \mathbb{R}^+ $. The weight under T-duality is
$w_1 - w_2$.  \label{qD=9}}
\end{table}

There are a doublet and a singlet of 1-forms, and what eq.
\eqref{D=9D0F0F0} shows is that the RR-form always belongs to the
doublet. Besides, no matter how we choose the RR-1-form within the
doublet, eq. \eqref{D=9D1F1} reveals that the RR-2-form must
correspond to the same component.\footnote{In comparing the doublet
of eq. \eqref{D=9D0F0F0} with the doublet of eq. \eqref{D=9D1F1} one
should keep in mind that the $\text{SL}(2, \mathbb{R})$ indices are
raised and lowered by means of the epsilon symbol.} A similar
analysis can be performed for the higher rank branes. The
decomposition of the 1-forms in terms of RR and F fields allows us
to identify the $q$ and $\tilde{q}$ charges as
  \begin{equation}
  C_1 = \tilde{q}^\alpha A_{1, \alpha}\quad  \quad B_1 = q^\alpha A_{1
  ,\alpha} \quad \quad B_1^\prime = q A_1 \quad .
  \end{equation}
This breaks $\text{SL}(2, \mathbb{R})$ to the subgroup
$\text{SO}(1,1)$, which is isomorphic to $\mathbb{R}^+$. We denote
with $w_1$ the charge associated with this $\mathbb{R}^+$, while the
charge associated to the original $\mathbb{R}^+$ is denoted with
$w_2$. The actual T-duality group is a linear combination of these two $\mathbb{R}^+$-factors
such that the weight $w$ under the T-duality group is given by $w=w_1-w_2$.
We summarise the decompositions of all the fields in
Table~\ref{qD=9}.

By analysing the table, one can determine the charges for all the
other fields. For the 2-forms $A_{2, \alpha}$ one has
  \begin{equation}
  (-1,1 )  : \ \tilde{q}^\alpha q   \qquad (1,1)  : \
q^\alpha q  \quad ,
\end{equation}
for the 3-form $A_3$ the charge is
  \begin{equation}
  (0,1)  : \ \epsilon_{\a\b} \tilde{q}^\alpha q^\beta q
  \end{equation}
and for the 4-form $A_4$ is
  \begin{equation}
  (0,2)  : \ \epsilon_{\a\b} \tilde{q}^\alpha q^\beta q^2 \quad
.
  \end{equation}
The expression for the RR and Fundamental 2-forms is thus
  \begin{equation}
  C_2 = \tilde{q}^\alpha q  A_{2 ,\alpha} \qquad B_2 = q^\alpha q
  A_{2 ,\alpha}
  \quad ,
  \end{equation}
and similarly for the higher rank forms.

Denoting with $n(\tilde{q})$, $n(q)$ and $n^\prime (q)$ the number
of times the charges $\tilde{q}^\alpha$, $q^\alpha$ and $q$
respectively occur in the decomposition of a given charge, one has
the relations
  \begin{equation}
  n(q) - n(\tilde{q}) = w_1 \quad \qquad n^\prime (q) = w_2 \quad .
  \label{D=9nwrelation}
  \end{equation}
Using this and the actual $\text{SL}(2, \mathbb{R})$ representation
to which each field belongs, the reader can identify the charges
corresponding to the higher rank fields. For instance, the 9-form in
the quadruplet is $A_{9 , \a\b\g}$, and its $(3,4)$ component is
projected by
   \begin{equation}
  (3,4)  : \ (\epsilon_{\a\b} \tilde{q}^\alpha q^\beta ) q^\gamma
q^\delta q^\epsilon q^4 \quad ,
\end{equation}
which is a RR field ($n(\tilde{q}) =1$), while for the 9-form in the
second doublet, $A_{9, \alpha}^\prime$, the component $(-1,2)$ is
projected by
  \begin{equation}
  (-1,2) : \ ( \epsilon_{\a\b} \tilde{q}^\a q^\b )^3 \tilde{q}^\gamma
q^2 \quad ,
\end{equation}
which has $n(\tilde{q}) = 4$. Finally, the relation between the
charges $w_1$ and $w_2$ of an $n$-form and the dilaton scaling of
the tension of the corresponding $(n-1)$-brane (if any) in the string
frame is
  \begin{equation}
  \alpha = \frac{1}{2} (w_1 + w_2 -n ) \quad ,
  \end{equation}
which is in agreement with \eqref{original} using
\eqref{D=9nwrelation} and
  \begin{equation}
 n = n(\tilde{q} ) + n(q) + n^\prime (q) \quad .
\end{equation}

For convenience, we now write down explicitly the gauge
transformations of the $C$ and $B$ fields. The two Fundamental
1-forms $B_1$ and $B_1^\prime$ transform as
  \begin{equation}
  \delta B_1 = d \Sigma_0 \qquad \delta B_1^\prime = d
\Sigma^\prime_0 \quad ,
  \end{equation}
and we denote their fields strengths as
  \begin{equation}
 H_2 = d B_1 \qquad H_2^\prime = d B_1^\prime \quad .
\end{equation}
The Fundamental 2-form $B_2$ transforms as
  \begin{equation}
  \delta B_2 = d \Sigma_1 - \frac{1}{2} ( \Sigma^\prime_0 H_2 +
\Sigma_0 H^\prime_2 ) \quad , \label{D=9gaugetransfofB2}
  \end{equation}
where the relative normalisation between the two terms in brackets
has been chosen for convenience (one can always change it by a field
redefinition of the form $B_2 \rightarrow B_2 + B_1 B_1^\prime$).
The gauge invariant field strength is
  \begin{equation}
  H_3 = d B_2 + \frac{1}{2} ( B_1^\prime H_2 + B_1 H_2^\prime )
\label{D=9H3dBdefinition}
  \end{equation}
and in the next section we will be needing the Bianchi identity
  \begin{equation}
  d H_3 = H_2^\prime H_2 \quad .
\end{equation}

From Table \ref{qD=9} one obtains that the gauge transformations of
the $C$ fields of even rank are
  \begin{equation}
  \delta C_{2n} = d \lambda_{2n-1} + H_3 \lambda_{2n-3} - H_2^\prime
\lambda_{2n-2} \quad , \label{D=9Ctransfeven}
  \end{equation}
while the gauge transformations of the $C$ fields of odd rank are
  \begin{equation}
  \d C_{2n+1} = d \lambda_{2n} + H_3 \lambda_{2n-2} - H_2
\lambda_{2n-1} \quad . \label{D=9Ctransfodd}
\end{equation}

\subsection*{D=8}
\begin{table}[!h]
\begin{center}
\begin{tabular}{|c|c||c|c|c|}
\hline \rule[-1mm]{0mm}{6mm} field & U repr & RR &  F & Rest\\
\hline \hline \rule[-1mm]{0mm}{6mm} 1-form & $({\bf
{\overline{3}}},{\bf 2}) $ &
$({\bf 1}, {\bf 2})_{-2}   $ & $({\bf 2},{\bf 2})_1$ &  \\
\hline \rule[-1mm]{0mm}{6mm} 2-form & $({\bf 3},{\bf 1})$ & $({\bf
2},{\bf 1})_{-1} $ &
$({\bf 1},{\bf 1})_{2}$ & \\
\hline \rule[-1mm]{0mm}{6mm} 3-form & $({\bf {1}},{\bf 2})$ & $({\bf 1},{\bf 2})_0$ & & \\
\hline \rule[-1mm]{0mm}{6mm} 4-form & $({\bf \overline{3}},{\bf 1})$
& $({\bf 2},{\bf 1})_1$ & & $({\bf 1},{\bf 1})_{-2}$\\
\hline \rule[-1mm]{0mm}{6mm} 5-form & $({\bf 3},{\bf 2})$ & $({\bf
1},{\bf 2})_2$ & & $({\bf 2},{\bf 2})_{-1}$ \\
\hline \rule[-1mm]{0mm}{6mm} 6-form & $({\bf 8},{\bf 1})$ & $({\bf
2},{\bf 1})_3$ & & $({\bf 2},{\bf 1})_{-3} + ({\bf 1},{\bf 1})_{0}+ ({\bf 3},{\bf 1})_{0}$ \\
\cline{2-5} \rule[-1mm]{0mm}{6mm}  & $({\bf 1},{\bf 3})$ &  & & $({\bf 1},{\bf 3})_{0}$\\
\hline \rule[-1mm]{0mm}{6mm} 7-form & $({\bf 6},{\bf 2})$ & $({\bf
1},{\bf 2})_4$ & & $({\bf 2},{\bf 2})_{1} + ({\bf 3},{\bf 2})_{-2}$ \\
\cline{2-5} \rule[-1mm]{0mm}{6mm}  & $({\bf \overline{3}},{\bf 2})$
&  & & $({\bf 2},{\bf 2})_{1}+
({\bf 1},{\bf 2})_{-2}$\\
\hline \rule[-1mm]{0mm}{6mm} 8-form & $({\bf 15},{\bf 1})$ & $({\bf
2},{\bf 1})_5$ & & $({\bf 4},{\bf 1})_{-1} + ({\bf 2},{\bf
1})_{-1}+({\bf 3},{\bf 1})_{-4} + ({\bf 3},{\bf 1})_{2}
+({\bf 1},{\bf 1})_{2}$ \\
\cline{2-5} \rule[-1mm]{0mm}{6mm}  & $({\bf {3}},{\bf 3})$ & & &
$({\bf 2},{\bf 3})_{-1}+
({\bf 1},{\bf 3})_{2}$\\
\cline{2-5} \rule[-1mm]{0mm}{6mm}  & $2 \times({\bf {3}},{\bf 1})$ &
& & $2 \times \left[ ({\bf 2},{\bf 1})_{-1}+
({\bf 1},{\bf 1})_{2} \right]$\\
 \hline
\end{tabular}
\end{center}
  \caption{\sl The decomposition of the $n$-form potentials of $D=8$ maximal
supergravity. The U-duality symmetry is $\text{SL}(3,\mathbb{R})
\times \text{SL}(2,\mathbb{R})$ and the T-duality is
$\text{SL}(2,\mathbb{R}) \times \text{SL}(2,\mathbb{R})$, while the
subscript denotes the $\mathbb{R}^+$-- charge $w$. \label{qD=8}}
\end{table}

\noindent We consider the 8-dimensional theory from the IIA
perspective, and we thus reduce the fields and branes in
\eqref{IIAassignments}. We first consider the 0-branes. Upon
reduction we obtain the following six 0-branes:
  \begin{eqnarray}
    & &  g_\m{}^\sharp \rightarrow {\rm D0} \hskip 1truecm \ g_\m{}^9 \ \ \rightarrow {\rm F0 }\hskip
1truecm  \  g_\m{}^8
   \rightarrow {\rm F0} \hskip 1truecm
      \nonumber \\[.2truecm]
    & & A_{\m 89} \rightarrow  {\rm D0} \hskip 1truecm
    A_{\m 8 \sharp} \rightarrow  {\rm F0} \hskip 1truecm
     A_{\m 9 \sharp} \rightarrow  {\rm F0
}    \quad .
    \end{eqnarray}
Using the $\text{SL}(3,\mathbb{R})$ epsilon symbol
$\epsilon^{89\sharp} =1$ we can write the last line as
  \begin{eqnarray}
    & & A_{\m}{}^\sharp \rightarrow  {\rm D0} \hskip 1truecm
     A_{\m }{}^9 \rightarrow {\rm  F0} \hskip 1truecm
     A_{\m }{}^8 \rightarrow {\rm F0}
    \quad .
    \end{eqnarray}
We therefore end up with two triplets $(M=8,9,\sharp)$ in the ${\overline {\bf 3}}$ representation
of $\text{SL}(3,\mathbb{R})$:
\begin{eqnarray}
A_{\mu, M1} &=& \left(A_\mu{}^8\,, A_\mu{}^9\,, A_\mu{}^\sharp\right) = {\rm (F0\,, F0\,, D0)}\,,\nonumber\\
[.2truecm] A_{\mu, M2} &=& \left(g_\mu{}^8\,, g_\mu{}^9\,,
g_\mu{}^\sharp\right)\ =\ {\rm (F0\,, F0\,, D0)} \,.
\label{twotriplets0branesD=8}
\end{eqnarray}
Together, the two triplets transform as a doublet $A_{\mu,
M\alpha}\, (\alpha=1,2)$, i.e.~as a $(\overline{\bf 3},{\bf 2})$
representation of the U-duality group.

Next, we consider the 1-branes. We have
  \begin{eqnarray}
  & & A_{\m\n \sharp} \rightarrow  {\rm F1}\,,\hskip 1truecm
  A_{\m\n 9 } \rightarrow {\rm D1} \,,\hskip 1truecm
 A_{\m\n 8 } \rightarrow {\rm D1} \quad .
 \end{eqnarray}
This forms a single triplet in the $({\bf 3}, {\bf 1})$
representation of the U-duality group:
\begin{equation}
A_{\mu\nu}^M = \left(A_{\m\n 8}\,, A_{\m\n 9}\,,
A_{\m\n\sharp}\right) = {\rm (D1\,, D1\,, F1)}\,.
\label{D1D1F1inD=8}
\end{equation}

Finally, we consider the 2-branes. These are
  \begin{eqnarray}
  & & A_{\m\n\r} \rightarrow {\rm D2}\,,\hskip 2truecm
  A_{\m\n\r 8 9 \sharp } \rightarrow {\rm D2} \,. \label{D2D2inD=8}
  \end{eqnarray}
The two fields are $\text{SL}(3,\mathbb{R})$ singlets, which form an
$\text{SL}(2,\mathbb{R})$ doublet $A_{3, \alpha}$.

By looking at equations \eqref{twotriplets0branesD=8},
\eqref{D1D1F1inD=8} and \eqref{D2D2inD=8} one can see that there is
an $\text{SL}(2,\mathbb{R})$ inside $\text{SL}(3,\mathbb{R})$ which
leaves the D0-branes and the D2-branes invariant and transforms
covariantly the D1-branes (in our choice this is the
$\text{SL}(2,\mathbb{R})$ that rotates the first two components of
the triplet). It is also evident that the D0-branes and the
D2-branes transform covariantly with respect to the other
$\text{SL}(2,\mathbb{R})$, while the D1-branes are invariant. The
T-duality is thus $\text{SL}(2,\mathbb{R}) \times
\text{SL}(2,\mathbb{R})$. In Table \ref{qD=8} we give the
decomposition of all the fields in terms of the T-duality group. We
also give in the table the corresponding $\mathbb{R}^+$-- charge as a
subscript. The conventions for this charge in this case and in all
the lower dimensional ones are taken from \cite{Slansky}.

The table allows us to identify the $q$ charges. In particular, from
eq. \eqref{twotriplets0branesD=8} we see that the D0-branes are a
doublet of the U-duality $SL(2,\mathbb{R})$, and we thus write
  \begin{equation}
  C_{1,\alpha} = {\tilde q}^M A_{1,M\alpha}\,,\hskip 2truecm B_{1,\alpha\dot\alpha} =
  q^M_{\dot\alpha} A_{1,M\alpha}\, ,
  \end{equation}
where $\dot{\alpha}$ denotes the doublet of the
$\text{SL}(2,\mathbb{R})$ inside $\text{SL}(3,\mathbb{R})$. All
other charges projecting the higher rank forms can be uniquely
expressed as products of these basic charges following the general
analysis at the beginning of this section. Denoting with
$n(\tilde{q})$ and $n(q)$ the number of times the charges
$\tilde{q}^M$ and $q^M_{\dot{\a}}$ occur in the decomposition of a
given field, the $\mathbb{R}^+$-- charges of the n-form fields ($n =
n(\tilde{q})+ n(q)$) are related to  these numbers by
\begin{equation}
w  = -2 n(\tilde{q}) + n(q) \quad .
\end{equation}
In case the form is associated to a brane, the dilaton scaling of
the tension of a D-$(n-1)$-brane is given in terms $w$ as
\begin{equation}
g_s^\alpha\,,\hskip 2truecm \alpha = -\frac{1}{3}(n-w)\,.
\end{equation}

It is instructive to identify all the $q$ charges for the higher
rank fields applying to the eight-dimensional case the general
analysis at the beginning of this section. For the two forms $A_2^M$
eq. \eqref{expressionsforqtwos} becomes
  \begin{equation}
  ({\bf 2,1})_{-1} : \ {\tilde q}^N q^P_{\dot\alpha}\epsilon_{MNP}
\quad \quad ({\bf 1,1})_{2} : \  q^N_{\dot\alpha}
q^P_{\dot\beta}\epsilon^{\dot\alpha\dot\beta}\epsilon_{MNP} \quad ,
\end{equation}
so that the RR and Fundamental 2-forms are given by
  \begin{equation}
  C_{2 , \dot\a} = {\tilde q}^N q^P_{\dot\alpha}\epsilon_{MNP}
  A_{2}^M
\quad \quad B_2 =  q^N_{\dot\alpha}
q^P_{\dot\beta}\epsilon^{\dot\alpha\dot\beta}\epsilon_{MNP}A_2^M
\quad .
\end{equation}
For the 3-forms $A_{3, \alpha}$  eq. \eqref{q3q1q1q1} gives
  \begin{equation}
 ({\bf 1,2})_{0} : \ {\tilde q}^M q^N_{\dot\alpha} q^P_{\dot\beta}\epsilon^{\dot\alpha\dot\beta}
\epsilon_{MNP} \quad .
\end{equation}
One can actually determine the charge associated to all the fields
in Table \ref{qD=8}, not only the RR fields. For instance, for the
5-forms $A_{5}^M{}_\alpha$, one gets the charge projecting on the RR
field, which is the $({\bf 1,2})_2$,
  \begin{equation}
  ({\bf 1,2})_{2} : \ {\tilde q}^N{q}^P_{\dot{\alpha}} q^Q_{\dot\beta} q^R_{\dot\gamma} q^S_{\dot\delta}
\e^{\dot\a \dot\b}
\epsilon^{\dot\gamma\dot\delta}\epsilon_{NMQ}\epsilon_{PRS} \quad ,
\end{equation}
which has $n(\tilde{q}) = 1 $ and $n(q) =4 $, and  the charge
projecting on the $({\bf 2,2})_{-1}$,
  \begin{equation}
  ({\bf 2,2})_{-1} : \ {\tilde q}^N{\tilde q}^P q^Q_{\dot\beta} q^R_{\dot\gamma} q^S_{\dot\delta}
\epsilon^{\dot\gamma\dot\delta}\epsilon_{NMQ}\epsilon_{PRS} \quad ,
\end{equation}
which has $n(\tilde{q}) = 2 $ and $n(q) =3 $. One can show that any
other structure made of $q$'s and $\tilde{q}$'s with $n(q) +
n(\tilde{q}) = 5$ projecting the 5-form vanishes identically.

As an example, we now  wish to consider the non-trivial conjugacy
classes to which  the D5-branes belong. For this purpose, it is much
easier to consider the truncation to the even form sector, that is
all the even-form fields whose gauge transformations are generated
by the gauge transformations of the 2-form. In that case the basic
charges are given by
\begin{equation}
{\tilde Q}_{M\dot\alpha}\,,\hskip 2truecm Q_M\,.
\end{equation}
These basic charges define the RR and Fundamental 2-forms as
\begin{equation}
C_{2,\dot\alpha} = {\tilde Q}_{M\dot\alpha} A_2^M \,,\hskip 2truecm
B_2 = Q_M A_2^M\,.
\end{equation}
The charges of the even-form fields that survive the truncation can
be expressed as products of these basic charges. In particular, for
the 4-form $A_{4 ,M}$ one gets
 \begin{equation}
 ({\bf 2,1})_1 : \ {\tilde Q}_{N\dot\alpha}Q_P\epsilon^{MNP} \qquad
({\bf 1,1})_{-2} : \ {\tilde Q}_{N\dot\alpha} {\tilde
Q}_{P\dot\beta}\epsilon^{\dot\alpha\dot\beta}\epsilon^{MNP} \quad ,
 \end{equation}
while for the 6-form $A_{6,P}{}^Q$ one gets\footnote{This 6-form is
the $({\bf 8,1})$. The other 6-form, the $({\bf 1,3})$, disappears
in the even-form truncation because its gauge transformations do not
talk to the gauge transformations of the lower rank even forms.}
  \begin{eqnarray}
  & & ({\bf 2,1 })_3 : \ {\tilde Q}_{M\dot\alpha}Q_N Q_Q\epsilon^{MNP}
\quad \quad ({\bf 2,1 })_{-3} : \ {\tilde
Q}_{M\dot\alpha}{\tilde Q}_{N\dot\beta} {\tilde Q}_{Q\dot\gamma}\epsilon^{\dot\beta\dot\gamma}\epsilon^{MNP} \nonumber \\
 & & ({\bf 3,1 })_0 : \ {\tilde Q}_{Q(\dot\alpha}
{\tilde Q}_{M\dot\beta )}Q_N\epsilon^{MNP}\quad  \quad ({\bf 1,1
})_0 : \ {\tilde Q}_{M\dot\alpha}{\tilde
Q}_{N\dot\beta}Q_Q\epsilon^{\dot\alpha\dot\beta}\epsilon^{MNP}
\end{eqnarray}
The $\mathbb{R}^+$-- charges are related to $n(\tilde{Q})$ and $n(Q)$
as
\begin{equation}
w = 2 n(Q) - n(\tilde{Q}) \,.
\end{equation}

Similar to the D7-branes in Type IIB string theory, we may now ask
which of the eight 5-branes can be reached by an $\text{SL}(3,
\mathbb{R})$ rotation of the two D5-branes, which correspond to the
$({\bf 2,1 })_3 $ fields. Consider the three independent
three-vectors
\begin{equation}
{\tilde Q}_{M\dot 1}\,,\hskip 1truecm {\tilde Q}_{M\dot 2}\,, \hskip
1truecm Q_M\,.
\end{equation}
We observe that in the expression for the charge of the D-brane two
of the three vectors are the same. This means that by an
$\text{SL}(3,\mathbb{R})$ rotation one can reach only those branes
whose expressions for the charge also contains two vectors that are
the same. This applies to the two ${\bf (2,1)}_{-3}$ branes and to two
of the three ${\bf (3,1)}_0$ branes. This implies that the D5-branes
describe a non-linear sixplet embedded into the octoplet. The two
remaining branes have charges such that the three $Q$'s are all
different. One may verify that for those branes one cannot write
down a gauge-invariant WZ-term.

The same conclusion can be reached considering the 0-brane charges
$q$ and $\tilde{q}$. To summarise, the $q$'s and $\tilde{q}$'s
select covariantly an $\text{SL}(2, \mathbb{R})$ inside
$\text{SL}(3, \mathbb{R})$, and rotating these charges under
$\text{SL}(3, \mathbb{R})$ corresponds to choosing a different
embedding. All the components of a representation of $\text{SL}(3,
\mathbb{R})$ that can be reached this way form a conjugacy class.
This applies to any dimension.

\subsection*{D=7}

 \begin{table}[!h]
\begin{center}
\begin{tabular}{|c|c||c|c|c|}
\hline \rule[-1mm]{0mm}{6mm} field & U repr & RR &  F & Rest\\
\hline \hline \rule[-1mm]{0mm}{6mm} 1-form & ${\bf \overline{10}}$ &
${\bf \overline{4}}_{-3}$ & ${\bf {6}}_2$ & \\
\hline \rule[-1mm]{0mm}{6mm} 2-form & ${\bf 5}$ & ${\bf 4}_{-1}$ &
${\bf 1}_{4}$ & \\
\hline \rule[-1mm]{0mm}{6mm} 3-form & ${\bf \overline{5}}$ & ${\bf \overline{4}}_{1}$ & & ${\bf 1}_{-4}$\\
\hline \rule[-1mm]{0mm}{6mm} 4-form & ${\bf 10}$ & ${\bf {4}}_{3}$ & & ${\bf 6}_{-2}$\\
\hline \rule[-1mm]{0mm}{6mm} 5-form & ${\bf 24}$ & ${\bf \overline{4}}_5$ & & ${\bf 15}_0 + {\bf 4}_{-5} + {\bf 1}_0$\\
\hline \rule[-1mm]{0mm}{6mm} 6-form & ${\bf \overline{40}}$ & ${\bf {4}}_7$ & & ${\bf \overline{20}}_{-3} + {\bf 10}_{2} + {\bf 6}_2$\\
\cline{2-5} \rule[-1mm]{0mm}{6mm}  & ${\bf \overline{15}}$ &  & & ${\bf \overline{10}}_2 + {\bf \overline{4}}_{-3} + {\bf 1}_{-8}$\\
\hline \rule[-1mm]{0mm}{6mm} 7-form & ${\bf {70}}$ & ${\bf
\overline{4}}_9$ & & ${\bf {36}}_{-1}
+ {\bf 15}_{4} + {\bf 10}_{-6} + {\bf 4}_{-1} + {\bf 1}_{4}$\\
\cline{2-5} \rule[-1mm]{0mm}{6mm}  & ${\bf {45}}$ &  & & ${\bf 20}_{-1} + {\bf {15}}_{4} +{\bf 6}_{-6} + {\bf 4}_{-1}$\\
\cline{2-5} \rule[-1mm]{0mm}{6mm}  & ${\bf {5}}$ &  & & ${\bf 4}_{-1} + {\bf 1}_4$\\
 \hline
\end{tabular}
\end{center}
  \caption{\sl The decomposition of the $n$-form potentials of $D=7$ maximal
supergravity. The U-duality group is $\text{SL}(5, \mathbb{R})$ and
the T-duality group is $\text{SL}(4, \mathbb{R})$. We denote as a
subscript the $\mathbb{R}^+$-- charge (notation from \cite{Slansky}).
\label{qD=7}}
\end{table}
\noindent We now consider the seven-dimensional case. From eq.
\eqref{IIAassignments} one finds that the 1-form fields associated
to the D0-branes are
  \begin{equation}
  g_\m{}^\sharp \qquad  A_{\m 89} \qquad A_{\m 97}\qquad A_{\m 87}
   \quad .
  \end{equation}
There is a manifest $\text{SL}(4, \mathbb{R})$ symmetry associated
to the torus, and using the corresponding epsilon symbol these
fields can be written as
  \begin{equation}
  A_{\m , 987} \qquad A_{\m}^{\sharp 7} \qquad  A_{\m}^{\sharp 8} \qquad  A_{\m}^{\sharp 9} \quad .
  \end{equation}
Using the $\text{SL}(5,\mathbb{R})$ symmetry enhancement, and
denoting with $6'$ the extra index of $\text{SL}(5,\mathbb{R})$, one
can now use the corresponding epsilon symbol on the first vector to
obtain
  \begin{equation}
  A_\m{}^{\sharp  6^\prime} \qquad A_{\m}^{\sharp 7} \qquad  A_{\m}^{\sharp 8} \qquad  A_{\m}^{\sharp 9} \quad .
 \end{equation}
Thus the D-branes correspond to the components of the vector $A_{\m
, MN}$ in the ${\bf \overline{10}}$ of $\text{SL}(5,\mathbb{R})$
with one index $\sharp$. The $\text{SL}(4,\mathbb{R})$ which
transforms the other four indices is the T-duality. Therefore the
D-branes belong to the ${\bf \overline{4}}$ of the T-duality group,
while the remaining 1-forms have $\text{SL}(5,\mathbb{R})$ indices
different from $\sharp$, which identifies the ${\bf 6}$ of
$\text{SL}(4,\mathbb{R})$. The corresponding branes are Fundamental.
In order to realise this covariantly, we therefore introduce the
charges $\tilde{q}^{MN}_a$ and $q^{MN}_{ab}$ that identify the RR
and Fundamental 1-forms:
 \begin{equation}
  C_{1 ,a}  = \tilde{q}^{MN}_a A_{1, MN} \qquad B_{1, ab} =
q^{MN}_{ab} A_{1, MN} \quad . \end{equation}

The 2-forms that result from the reduction of \eqref{IIAassignments}
are
  \begin{equation}
  A_{\m\n \sharp} \qquad   A_{\m\n 9} \qquad   A_{\m\n 8} \qquad   A_{\m\n
7} \qquad  A_{\m\n 789\sharp} \quad ,
  \end{equation}
and the last component can be rewritten as $A_{\m\n 6'}$, which
makes the $\bf 5$ of $\text{SL}( 5,\mathbb{R})$. From
\eqref{IIAassignments} it follows that the first component $ A_{\m\n
\sharp}$ corresponds to the Fundamental string, while the other
components are associated to D1-branes. This is in agreement with
eq. \eqref{expressionsforqtwos}, which in this seven-dimensional
case says that the 2-forms $A_2^M$ are projected on the RR 2-forms
in the $\bf 4$ of $\text{SL}(4, \mathbb{R})$  by
  \begin{equation}
  C_{2}^{ a} = \tilde{q}^{NP}_b q^{QR}_{cd} \epsilon^{abcd}
  \epsilon_{MNPQR} A_{2}^M
  \end{equation}
and on the Fundamental 2-form by
  \begin{equation}
B_2 = q^{NP}_{ab} q^{QR}_{cd} \epsilon^{abcd} \epsilon_{MNPQR} A_2^M
\quad .
\end{equation}

The decomposition of all the fields under T-duality is given in
Table \ref{qD=7}. The relation between the $\mathbb{R}^+$-- charge $w$
and the $q$'s is
  \begin{equation}
  w = -3 n(\tilde{q}) + 2 n(q) \quad .
  \end{equation}
From this one can determine all the $\mathbb{R}^+$-- charges of the
 various T-duality representations given in Table \ref{qD=7}.

\subsection*{D=6}
 \begin{table}[!h]
\begin{center}
\begin{tabular}{|c|c||c|c|c|}
\hline \rule[-1mm]{0mm}{6mm} field & U repr & RR &  F & Rest\\
\hline \hline \rule[-1mm]{0mm}{6mm} 1-form & ${\bf {16}}$ &
$({\bf 8_{\rm S}})_{-1}$ & $({\bf 8_{\rm C}})_{1}$ &  \\
\hline \rule[-1mm]{0mm}{6mm} 2-form & ${\bf {10}}$ & $({\bf 8_{\rm
V}})_{0}$ &
${\bf 1}_{2}$ & ${\bf 1}_{-2}$\\
\hline \rule[-1mm]{0mm}{6mm} 3-form & ${\bf \overline{16}}$ & $({\bf 8_{\rm S}})_{1}$ & & $({\bf 8_{\rm C}})_{-1}$\\
\hline \rule[-1mm]{0mm}{6mm} 4-form & ${\bf {45}}$ & $({\bf 8_{\rm V}})_{2}$ & & $({\bf 8_{\rm V}})_{-2} + {\bf 28}_0 + {\bf 1}_0$\\
\hline \rule[-1mm]{0mm}{6mm} 5-form & ${\bf {144}}$ & $({\bf 8_{\rm
S}})_{3}$ & & $({\bf 8_{\rm C}})_{1} + ({\bf 8_{\rm V}})_{-1} +
({\bf 8_{\rm C}})_{-3} + ({\bf 56_{\rm V}})_{- 1 } + ({\bf 56_{\rm
C}})_{1}$\\
\hline \rule[-1mm]{0mm}{6mm} 6-form & ${\bf {320}}$ & $({\bf
{8}_{\rm V}})_{4}$ & & $({\bf 8_{\rm V}})_{-4} + 2\times ({\bf
8_{\rm V}})_{0} + ({\bf 35_{\rm V}})_{2} + ({\bf 35_{\rm V}})_{- 2 }
+ ({\bf 160_{\rm V}})_{0}
$ \\
\rule[-1mm]{0mm}{6mm} && & &  $+ {\bf {28}}_{2}+ {\bf 28}_{-2} +
{\bf {1}}_{2}+ {\bf 1}_{-2}
$\\
\cline{2-5} \rule[-1mm]{0mm}{6mm}  & ${\bf \overline{126}}$ &  & & $
({\bf 35_{\rm S}})_{2} + ({\bf 35_{\rm C}})_{-2 } + ({\bf 56_{\rm
V}})_0 $\\
\cline{2-5} \rule[-1mm]{0mm}{6mm}  & ${\bf 10}$ &  & & $ ({\bf
8_{\rm V}})_{0} + {\bf 1}_{2 } + {\bf 1}_{-2} $\\
 \hline
\end{tabular}
\end{center}
  \caption{\sl The decomposition of the $n$-form potentials of $D=6$ maximal
supergravity. The U-duality is $\text{SO}(5,5)$, while the T-duality
is $\text{SO}(4,4)$. \label{qD=6}}
\end{table}

\noindent In six dimensions the U-duality group is SO(5,5) which is
decomposed under the T-duality group SO(4,4). This is given in Table
\ref{qD=6}. We follow the group theory conventions of
\cite{Slansky}, and therefore the RR 2-forms belong to the ${\bf
8_{\rm V}}$ of SO(4,4). This is not in contradiction with the
general case, in which all RR fields are in the spinor
representations of the T-duality group, because of triality of
SO(4,4).

As usual, we denote with $w$ the $\mathbb{R}^+$-- charge. For the
cases in which the $n$-form  can be associated to a brane, the
corresponding tension scales in the string frame as
  \begin{equation}
  \alpha = \frac{1}{2} ( w -n) \quad .
\end{equation}
Only one of the two singlets that arise in the decomposition of the
2-forms is a Fundamental string. The other singlet corresponds to a
string scaling like $g_s^{-2}$, which is the magnetic dual of the
Fundamental string.

\subsection*{D=5}

\noindent In five dimensions the 1-forms, which belong to the ${\bf
27}$ of $\text{E}_6$, decompose into ${\bf 16+ 10+1}$ under
T-duality. The decomposition of all the fields is given in Table
\ref{qD=5}, where the subscript denotes the $\mathbb{R}^+$-- charge
$w$. For the cases in which an $(n-1)$-brane is associated to the
$n$-form, the corresponding tension scales in the string frame as
  \begin{equation}
  \alpha = -\frac{1}{3} ( w +2 n) \quad .
\end{equation}
This shows that the 1-form singlet is not a Fundamental particle.
This is the highest dimension in which the 1-forms are not completely
decomposed into RR and Fundamental fields.
\begin{table}[!h]
\begin{center}
\begin{tabular}{|c|c||c|c|c|}
\hline \rule[-1mm]{0mm}{6mm} field & U repr & RR &  F & Rest\\
\hline \hline \rule[-1mm]{0mm}{6mm} 1-form & ${\bf {27}}$ &
${\bf {16}}_{1}$ & ${\bf {10}}_{-2}$ & ${\bf 1}_{4}$ \\
\hline \rule[-1mm]{0mm}{6mm} 2-form & ${\bf \overline{27}}$ & ${\bf
\overline{16}}_{-1}$ &
${\bf 1}_{-4}$ & ${\bf 10}_2$\\
\hline \rule[-1mm]{0mm}{6mm} 3-form & ${\bf {78}}$ & ${\bf {16}}_{-3}$ & & ${\bf \overline{16}}_{3}+ {\bf 45}_0 + {\bf 1}_0$\\
\hline \rule[-1mm]{0mm}{6mm} 4-form & ${\bf 351}$ & ${\bf
\overline{16}}_{-5}$ & & ${\bf 16}_{1} + {\bf 45}_4
+ {\bf 120}_{-2} + {\bf 144}_1 +{ \bf 10}_{-2}$\\
\hline \rule[-1mm]{0mm}{6mm} 5-form & ${\bf \overline{1728}}$ &
${\bf {16}}_{-7}$ & & ${\bf 1}_{-4} + {\bf 10}_2 + 2 \times{\bf
\overline{16}}_{-1} + {\bf 45}_{-4} + {\bf 120}_2 + {\bf
\overline{126}}_2 $ \\
\rule[-1mm]{0mm}{6mm} && & &  $+ {\bf \overline{144}}_{-1}+ {\bf
144}_5 + {\bf {210}}_{-4}+ {\bf 320}_2
+ {\bf \overline{560}}_{-1}$\\
\cline{2-5} \rule[-1mm]{0mm}{6mm}  & ${\bf \overline{27}}$ &  & & ${\bf \overline{16}}_{-1} + {\bf 10}_2 + {\bf 1}_{-4}$\\
 \hline
\end{tabular}
\end{center}
  \caption{\sl The decomposition of the $n$-form potentials of $D=5$ maximal
supergravity. The U-duality is $\text{E}_6$ and the T-duality is
SO(5,5). \label{qD=5}}
\end{table}

\subsection*{D=4}

We first consider the 1-forms, see Table \ref{qD=4}. The RR 1-forms
occur as a 32-dimensional spinor representation of the T-duality
group $\text{SO}(6,6)$, as expected. There are 24 remaining 1-forms,
half of them correspond to Fundamental $0$-branes, the other half
correspond to Solitonic 0-branes with dilaton coupling $1/g_s^2$. To
pick out the Fundamental fields one must therefore   undo the
symmetry enhancement by making the decomposition \eqref{furtherD=4}.
Under this decomposition the 1-forms branch as

\begin{equation}
{\bf (32,1)}\ \rightarrow \ {\bf 32}_0\,,\hskip 2truecm {\bf (12,2)}\ \rightarrow \ {\bf 12}_{-1} + {\bf 12}_1\,,
\end{equation}
where the sub-index indicates the weight $w$ under $\mathbb{R}^+$.
One next uses the rule
that the brane tension corresponding to an $n$-form scales as
\begin{equation}
g_s^\alpha\,,\hskip 2truecm \alpha = -(n+w)\,.
\end{equation}
This confirms that the ${\bf 32}_0$ charges describe 32 D$0$-branes. We furthermore deduce  that the
${\bf 12}_{-1}$ charges describe 12 Fundamental 0--branes and that  the ${\bf 12}_1$
charges describe 12 Solitonic 0-branes.

Like in $D=5$ dimensions, the 1-forms are decomposed not only in RR
and Fundamental fields but also in Solitonic fields $D$. We
therefore introduce the three basic charges $({\tilde q}^M_a, q^M_A,
q'^M_A)$ and decompose the 1-forms as follows:

\begin{equation}
C_{1,a} = {\tilde q}^M_a A_{1,M}\,,\hskip 1truecm B_{1,A} = q^M_A A_{1,M}\,,\hskip 1truecm D_{1,A} = q'^M_A A_{1,M}\,.
\end{equation}
In terms of these charges the weight $w$ of an $n$-form is given by
\begin{equation}
w = -n(q)+n(q')\,.
\end{equation}

We next consider the 2-forms $A_{2 ,\alpha}$. Under the
decomposition \eqref{furtherD=4} the 2-forms in the RR column
branch according to
\begin{equation}
{\bf (32^\prime,2)}\ \rightarrow \ {\bf 32}^\prime_{-1} +{\bf 32}^\prime_1
\end{equation}
This shows that the ${\bf 32}^\prime_{-1}$  2-forms describe D1-branes but that the ${\bf 32}^\prime_1$  2-forms describe exotic objects with
$1/g_s^3$ dilaton coupling. Their charges are given by

\begin{equation}
{\bf 32}^\prime_{-1}\,:\quad {\tilde q}^M_aq^N_A(\Gamma^A)_{\dot a}{}^a D_{MN}^\alpha\,,\hskip 2truecm
{\bf 32}^\prime_1\,:\quad {\tilde q}^M_aq'^N_A(\Gamma^A)_{\dot a}{}^a D_{MN}^\alpha\,.
\end{equation}
Similarly, the 2-forms in the F column decompose according to
\begin{equation}
{\bf (1,3)}\ \rightarrow {\bf 1}_{-2} + {\bf 1}_0 + {\bf 1}_2\,.
\end{equation}
They describe a Fundamental string, a Solitonic string and an exotic object with $1/g_s^4$ coupling, respectively.
Their charges are given by

\begin{equation}
 {\bf 1}_{-2}\,:\quad q^M_A q^N_B\eta^{AB} D_{MN}^\alpha\,,\hskip 1truecm
{\bf 1}_0\,:\quad q^M_A q'^N_B\eta^{AB} D_{MN}^\alpha\,,\hskip 1truecm
{\bf 1}_2\,:\quad q'^M_A q'^N_B\eta^{AB} D_{MN}^\alpha\,.\hskip 1truecm
\end{equation}
Finally, the remaining 2-forms in the Rest column decompose according to
\begin{equation}
{\bf (66,1)}\ \rightarrow \ {\bf 66}_0\,.
\end{equation}
They correspond to Solitonic strings with charges given by
\begin{equation}
{\bf 66}_0\,:\quad q^M_{[A} q'^N_{B]}D_{MN}^\alpha\,.
\end{equation}
Similarly, one can discuss the higher $n$-forms.

\begin{table}[!h]
\begin{center}
\begin{tabular}{|c|c||c|c|c|}
\hline \rule[-1mm]{0mm}{6mm} field & U repr & RR &  F & Rest\\
\hline \hline \rule[-1mm]{0mm}{6mm} 1-form & ${\bf {56}}$ &
${\bf {(32,1)}}$ & ${\bf {(12,2)}}$ &  \\
\hline \rule[-1mm]{0mm}{6mm} 2-form & ${\bf 133}$ & ${(\bf
32^\prime, 2)}$ &
${\bf (1,3)}$ & ${\bf (66,1)}$\\
\hline \rule[-1mm]{0mm}{6mm} 3-form & ${\bf {912}}$ & ${\bf (32,3)}$ & & ${\bf (12,2)}+ {\bf (352,1)} + {\bf (220,2)}$\\
\hline \rule[-1mm]{0mm}{6mm} 4-form & ${\bf {8645}}$ &
${\bf (32^\prime, 4) }$ & & ${\bf (2079,1)} + {\bf (1728,2)} + {\bf (495,3)} + {\bf (462,1)}$  \\
\rule[-1mm]{0mm}{6mm} & & & & $+ {\bf (352,2)}+ {\bf (66,3)} + {\bf (66,1)} + {\bf (32^\prime , 2)} + {\bf (1,3)}$ \\
\cline{2-5} \rule[-1mm]{0mm}{6mm}  & ${\bf {133}}$ &  & & ${\bf (32^\prime ,2 )} + {\bf (1,3)} + {\bf (66,1)}$\\
 \hline
\end{tabular}
\end{center}
  \caption{\sl The decomposition of the $n$-form potentials of $D=4$ maximal supergravity. The U-duality is
  $E_7$ and the T-duality is $\text{SO}(6,6)$ with symmetry enhancement to $\text{SO(6,6)} \times \text{SL}(2,
\mathbb{R})$. The $n$-forms in the RR and F columns do not only
contain the RR and Fundamental fields but also fields with different
dilaton couplings, see the text. \label{qD=4}}
\end{table}

%
%

\section{D-brane WZ Terms}

In this section we will derive the main result of this paper,
i.e.~the expression \eqref{WZtermD<10} for the general D-brane WZ
term in $3 \le D\le 10$ dimensions. Our starting point is the
U-duality covariant expressions for the RR $C$ fields in terms of
the covariant $A$ fields, derived in the previous section. These RR
fields have the important property that  under gauge transformations
they transform only into themselves, see
eq.~\eqref{RRgaugetransfcompact}. They form the building blocks for
our construction of a gauge-invariant WZ term.

It is well-known that in ten dimensions it is not possible to
construct a gauge-invariant WZ term using the RR fields $C$ alone.
The reason is that the $C$ fields not only transform into a total
derivative but also into a term containing the Fundamental curvature
$H_3$ which has to be cancelled. Trying something of the form
$e^{X_2}C$ with $dX_2=H_3$ would solve the problem. Taking $X_2=B_2$
is not allowed since $X_2$ has to be gauge-invariant by itself. This
is the reason that we need to introduce a BI vector such that $B_2$
can be interpreted as a term inside the gauge-invariant curvature of
the BI vector. We therefore take $X_2={\cal F}_2=dV_1+B_2$.

In $D<10$ dimensions a similar reasoning works except that the $C$
fields now not only transform to $H_3$ but also to $H_{2,A}$, the
curvature of the Fundamental 1-forms $B_{1,A}$. This suggests that
we need to introduce not only a BI vector $V_1$ but also $2(10-D)$
worldvolume scalars $V_{0,A}$ with corresponding gauge invariant
field strengths given by
  \begin{equation}
  {\cal F}_{1 ,A} = d V_{0 ,A} + B_{1, A}.
  \end{equation}
This is gauge-invariant provided that the worldvolume scalars
transforms as
\begin{equation}
  \d V_{0 ,A} = - \Sigma_{0, A}\,.
  \end{equation}
Furthermore, we need to adapt the definition of ${\cal F}_2$ since
$B_2$ also transforms under $\Sigma_{0,A}$, see
eq.~\eqref{NSgaugetransfs}. The following expression is
gauge-invariant
  \begin{equation}
  {\cal F}_2 = d V_1 + B_2 - V_{0 ,A} H_{2 ,B} \eta^{AB} \quad ,
  \end{equation}
provided that $V_1$ transforms  under gauge transformations as
  \begin{equation}
  \d V_1 = - \Sigma_1 \quad .
  \end{equation}

It is instructive to compare the above with the expected number of
bosonic worldvolume degrees of freedom. In general a D$p$-brane has
8 bosonic worldvolume degrees of freedom. For instance, in ten
dimensions a D$p$-brane has $(p-1)$ d.o.f represented by the BI
vector and $(10-p-1)$ d.o.f. represented by the worldvolume scalars,
after fixing the $(p+1)$ worldvolume reparametrisations, i.e.

\begin{equation}
(p-1) + (10-p-1)= 8\,.
\end{equation}
In $D<10$ dimensions we not only have the $(p-1)$ d.o.f. represented
by the BI vector and the $D-p-1$ d.o.f. of the embedding scalars but
also the $(10-D)$ d.o.f. represented by the wrapping of the
Fundamental string around each of the $(10-D)$ compactified
dimensions. This leads to the same total number of d.o.f. as in
$D=10$:

\begin{equation}
(p-1) +(D-p-1) + (10-D) = 8\,.
\end{equation}
We find that the WZ term contains twice as much extra scalars than
expected, i.e. $2(10-D)$ instead of $(10-D)$. We will comment about
this in the conclusions section.

Given these ingredients, we can now write a compact expression for
the gauge-invariant WZ term for any D-brane in any dimension. The
result is given in \eqref{WZtermD<10} which we repeat here:
  \begin{equation}\label{ansatz}
  {\cal L}_{\text{WZ}}(D<10) = e^{{\cal F}_2 } e^{{\cal F}_{1 ,A}
  \Gamma^A} C \quad .
  \end{equation}
Like in the previous section we denote with $C$ the sum of all the
RR potentials. The $D=9$ case is a bit special in the sense that the
combination  ${\cal F}_{1 ,A} \Gamma^A$ only contains the self-dual
(anti-self-dual) part of ${\cal F}_{1,A}$ when projected on the even
(odd) form sector. These 1-forms contain the $(1,0)$ and $(0,1)$
1-forms in Table \ref{qD=9}. It is also  easier to explicitly write
out the $\Gamma^A$ matrices for this case. We therefore treat this
case separately below, see \eqref{special9D} for an expression for
the WZ term in that case.

To proof that the WZ term \eqref{ansatz} is
gauge-invariant we need the Bianchi identities of the worldvolume
curvatures:

\begin{eqnarray}
   & & d {\cal F}_{1 , A} = H_{2, A}\,,
   \nonumber \\[.1truecm]
   & & d {\cal F}_2 = H_3 - {\cal F}_{1, A} H_{2 , B} \eta^{AB}
   \label{curlofcalF}
   \quad .
   \end{eqnarray}
The proof of gauge-invariance is now remarkably simple.
 Given that the ${\cal F}$'s are gauge invariant, we
have
  \begin{equation}\label{variation}
  \delta {\cal L}_{\text{WZ}}(D<10) = e^{{\cal F}_2 } e^{{\cal F}_{1
  ,A} \Gamma^A}
  ( d \lambda + H_3 \lambda - H_{2 , B} \Gamma^B \lambda ) \quad .
  \end{equation}
The $H_3 $ term cancels up to a total derivative precisely like in
ten dimensions by using the second Bianchi identity of
\eqref{curlofcalF}. We are now left with the following three  terms,
leaving out an overall $e^{{\cal F}_2}$ factor and the gauge
parameter $\lambda$:
  \begin{equation}\label{three}
  e^{{\cal F}_{1 ,A}\Gamma^A} {\cal F}_{1, B} H_{2, C} \eta^{BC} - d   e^{-{\cal F}_{1
  ,A} \Gamma^A}
  -   e^{{\cal F}_{1 ,A}\Gamma^A} H_{2 , B} \Gamma^B \quad .
  \end{equation}
The first term arises from the fact that we applied the second
Bianchi identity of \eqref{curlofcalF} when cancelling the $H_3$
term. The second term arises from partially differentiating the
exterior derivative in \eqref{variation} when it hits the $e^{{\cal
F}_{1,A} \Gamma^A}$ term. Finally, the third term is just the last
term of \eqref{variation}. To show that the three terms given in
\eqref{three} cancel amongst each other it is convenient to first
expand the exponential in the first and second term

\begin{equation}
e^{{\cal F}_{1,A}\Gamma^A} = \sum_{n=0}^{2(10-D)}\frac{1}{n^!}{\cal
F}_{1,A_1}\cdots {\cal F}_{1,A_n}\Gamma^{A_1\cdots A_n}
\end{equation}
and similarly expand the exponential in the second term. In the
second term one next uses the  first Bianchi identity of
\eqref{curlofcalF}. Now all terms are linear in $H_2$. After
expanding the exponentials the first and second term of
\eqref{three} are written as a sum of completely anti-symmetric
Gamma matrices. The third term becomes the sum of products of an
anti-symmetric Gamma matrix with a single $\Gamma_B$ matrix. Working
out this product leads to two types of terms containing a single
anti-symmetric Gamma matrix. In the first type $H_2$ is contracted
with one of the indices of the anti-symmetric Gamma matrix. Such
terms cancel against the second term of \eqref{three}. In the second
type $H_2$ is contracted with one of the ${\cal F}_1$ curvatures.
These terms cancel against the first term of \eqref{three}. This
completes the proof of gauge invariance of the WZ term
\eqref{ansatz}.

The form of the WZ term \eqref{ansatz} clearly suggests that, like
in $D=10$, the RR scalars $C_{0,{\dot a}}$ can be included too in
the expression although they are not needed for gauge-invariance.
Like in IIB supergravity all RR scalars are axionic, i.e.~the
$D$-dimensional maximal supergravity theory is invariant under
constant shifts of these scalars.

Given that in nine dimensions the T-duality group is abelian, we
consider this case explicitly although it does not differ from the
general analysis. We introduce the world-volume scalars $V_0 $ and
$V^\prime_0$, such that
  \begin{equation}
  {\cal F}_1 = d V_0 + B_1 \qquad {\cal F}^\prime_1 = d V_0^\prime +
B^\prime_1
  \end{equation}
are gauge invariant. From \eqref{D=9gaugetransfofB2} one also
defines
  \begin{equation}
  {\cal F}_2 = d V_1 + B_2 - \frac{1}{2} ( V_0^\prime H_2 + V_0
H^\prime_2 ) \quad , \end{equation} so that the Bianchi identities
are
  \begin{eqnarray}
  & & d {\cal F}_1 = H_2 \qquad \quad d {\cal F}_1^\prime =
H_2^\prime \nonumber \\
& & d {\cal F}_2 = H_3 - \frac{1}{2} ( {\cal F}_1^\prime H_2 + {\cal
F}_1 H_2^\prime ) \quad .
\end{eqnarray}
From the gauge transformations \eqref{D=9Ctransfeven} and
\eqref{D=9Ctransfodd} one can then write a gauge invariant WZ term
as the formal expression
  \begin{equation}
 {\cal L}_{\text{WZ}}(D=9) =  e^{{\cal F}_2} e^{{\cal F}_1 + {\cal F}_1^\prime} C \quad .\label{special9D}
 \end{equation}
It is understood here that when we expand $e^{{\cal F}_1 + {\cal
F}_1^\prime} $ we only take ${\cal F}_1$  acting on even forms and
${\cal F}_1^\prime$ acting on odd forms. That is
  \begin{equation}
  e^{{\cal F}_1 + {\cal F}_1^\prime} C = (1 + {\cal F}_1 + \tfrac{1}{2} {\cal F}_1^\prime {\cal F}_1 ) C_{\rm
even} +   (1 + {\cal F}_1^\prime + \tfrac{1}{2} {\cal F}_1 {\cal
F}_1^\prime ) C_{\rm odd} \quad.
  \end{equation}
Using this notation the proof of gauge-invariance is straightforward and will not be repeated here.
The alternating occurrence of ${\cal F}_1$ and ${\cal F}_1^\prime$
is easily explained if one considers the IIA and IIB origin of the
WZ term. Indeed, an even form in nine dimensions is unwrapped from
the IIB perspective and wrapped from the IIA perspective, and
vice-versa for an odd form. Any time there is a wrapped coordinate,
this corresponds to a world volume scalar appearing in the WZ term.
In particular the scalar $V_0$ can be seen as arising from a wrapped
world-volume vector $V_1$ in IIB, while the scalar $V_0^\prime$
arises as a wrapped $V_1$ in IIA.

\section{Conclusions}

In this paper we have constructed  gauge-invariant and U-duality
covariant expressions for Wess-Zumino terms corresponding to general
D$p$-branes ($0\le p \le D-1$) in arbitrary $3\le D\le 10$
dimensions. We did this in two steps. First, we considered the
target space background fields. In particular, we constructed
expressions for the RR potentials in terms of the U-duality
covariant fields. For this we introduced two types of charge vectors
that project the 1-forms onto the RR and Fundamental
1-forms.\,\footnote{Actually, for $D\le 5$ we need extra basic
charges to project onto Solitonic 1-forms as well.} We showed how,
for $D<10$,  the charges of all D$p$-branes with $p\ge1$ could be
expressed as products of these basic charges and we derived general
expressions for these higher-dimensional charges. The cases $D=3$
and $D=4$ required special attention due to symmetry enhancements
that take place in these dimensions. Since the extra symmetries put
the D-branes together with other objects into the same multiplet it
is natural, for the purposes of this paper, to undo theses symmetry
enhancements. We discussed the $D=4$ case in quite some detail. We
refrained from giving the formulae for the $D=3$ case as well but we
expect that it follows the same pattern we found in higher
dimensions.

In a second step we considered the worldvolume fields needed for the
construction of the WZ term. Since, for $D<10$, the WZ term for
general D-branes contains both even-form and odd-form potentials it
is clear that one needs also even-form and odd-form worldvolume
curvatures. We therefore introduced, besides the usual worldvolume
2-form curvature for the BI vector,  additional worldvolume 1-form
curvatures for the extra worldvolume scalars that correspond to the
compactified dimensions. These scalars are on top of the usual
embedding scalars. Using the expressions for the RR potentials and
the worldvolume curvatures it was then a relatively straightforward
task to construct a gauge-invariant and duality-covariant expression
for the  WZ term in $D<10$ dimensions.

The fact that for $D<10$ the charges of the higher-dimensional
branes can be expressed in terms of products of the D$0$-brane and
Fundamental 0-brane charges has important consequences for the
non-trivial conjugacy classes to which these D-branes may belong. As
an example, we analysed in detail the case of D5-branes in $D=8$
dimensions. To simplify matters we did this for the truncated case
of even forms only. Making use of the observation that the charge of
the standard D5-brane contains the product of two charge vectors
that are the same we showed that under a general U-duality
transformation they form a non-linear six-plet embedded into an
octo-plet of 5-branes. This is similar to the D7-branes in IIB
string theory which form a non-linear doublet inside a triplet of
7-branes. Other examples of non-trivial conjugacy classes may be
discussed similarly.

A noteworthy feature of the general WZ term is that it contains
twice as many extra scalars as compactified dimensions. One set of
scalars has a natural IIA origin, the other set has a natural IIB
origin. Together they transform as a vector of the T-duality group
$\text{SO}(10-D,10-D)$. This doubling of compactified dimensions is
typical for doubled geometry \cite{Hull:2004in,Hull:2006va,Hull:2007jy,Albertsson:2008gq} 
but now applied to
the worldvolume of the D-branes in  a curved background.
In the same way that in  dimensions lower than ten the Wess-Zumino term of the $D=10$
fundamental string, ${\cal L}_{\text{F1}} = B_2$, gets modified by ${\cal F}_{1,A}$ to \cite{Hull:2004in}

 \begin{equation}
{\cal L}_{\text{F1}}(D<10) =   B_2 + \eta^{AB} {\cal F}_{1, A} B_{1 ,B} \quad ,
  \end{equation}
which is invariant under the NS-NS gauge transformations \eqref{NSgaugetransfs},
we have derived that in $D<10$ dimensions the WZ term
of the $D=10$ D-branes, ${\cal L}_{\text{Dp}} = e^{{\cal F}_2} C$ gets modified 
by ${\cal F}_{1,A}$ to the expression \eqref{ansatz}.
In the case of the fundamental string a correct counting of the worldvolumne scalars is re-obtained by imposing
a self-duality condition on the scalars \cite{Hull:2004in}. 
The challenge will be  to see what kind of condition on the worldvolume scalars must be imposed, in the case of D-branes,  to 
obtain a  correct counting of the worldvolume degrees of freedom.
 Given that such a condition can be imposed for the fundamental string, we expect this to be a solvable
problem \cite{inpreparation2}.

There are two natural extensions of our work. One extension is to
consider the coupling of D-branes to maximal {\sl gauged}
supergravities. This requires the introduction of the so-called
embedding tensor \cite{Nicolai:2001sv,deWit:2002vt,deWit:2003hr}.
Like in the case of the coupling of the D2-brane to massive IIA
supergravity, we expect this embedding tensor to occur as the
coefficient of a worldvolume Chern-Simons term
\cite{Bergshoeff:1996cy}. More information about these extra
Chern-Simons terms can be obtained by considering generalized
Scherk-Schwarz reductions of the axionic RR scalars since these
reductions lead to Chern-Simons terms. We plan to come back to the
relation between the D-brane WZ terms and the embedding tensor in
the nearby future \cite{inpreparation1}.

A second extension is to construct a kappa-symmetric version of the
D-brane actions. This requires of course first to construct a
U-duality covariant  kinetic term for all worldvolume fields. This
has already been done for the $D=10$ D-branes, see
\cite{Bergshoeff:2006gs}, and a kappa-symmetric extension of the
$D=10$ D-brane actions has been constructed
\cite{Bergshoeff:2007ma}. 

Finally, we observe that lower-dimensional supergravities allow for
many potentials that correspond to exotic branes with unconventional
dilaton couplings $1/g_s^n$ \ $(n=3,4,5,\cdots)$. It is not clear
what the status of all these exotic branes is in string theory, and
if they exist et all, but if they do exist they might help in
finding a geometrical description of the individual degrees of
freedom of black holes and thereby explaining the entropy of these
black holes \cite{JdeBoer}.

\section*{Acknowledgements}

We thank C. Bachas, A. Marrani and N. Obers for useful discussions.
E.B. wishes to thank King's College London and F.R. wishes to thank
the University of Groningen for hospitality. The work of F.R. was
supported by the STFC rolling grant ST/G000/395/1.

\end{document}